\documentclass[11pt]{article}

\usepackage[utf8]{inputenc}
\usepackage{amsmath}
\usepackage{amssymb}
\usepackage{mathtools}
\usepackage{bm}
\usepackage{color}
\usepackage{units}
\usepackage{graphicx}
\usepackage{hyperref}
\usepackage[letterpaper, margin=1in]{geometry}
\usepackage{authblk}
\usepackage{booktabs}
\usepackage{multirow}
\usepackage{outlines}
\usepackage{filecontents}
\usepackage[version=3]{mhchem} % Formula subscripts using \ce{}
\usepackage{textcomp}
\usepackage{gensymb}
\usepackage[normalem]{ulem}
\usepackage{mhchem}
\usepackage[symbol]{footmisc}
\usepackage[labelfont=bf]{caption}

\newcommand{\ve}[1]{\boldsymbol{\mathbf{#1}}}
\newcommand{\kbnote}[1]{\textcolor{red}{#1}}

\newcommand{\RE}{\text{Re}}
\newcommand{\slow}{\text{slow}}
\newcommand{\fast}{\text{fast}}

\newcommand{\sh}{\text{sh}}

\usepackage[
backend=bibtex,
style=nature,
doi=false,
isbn=false,
url=false,
maxbibnames=99
]{biblatex}
\AtEveryBibitem{\clearfield{month}}
\AtEveryCitekey{\clearfield{month}}

\title{Patterns that persist: Heritable information in stochastic dynamics}

\author{Peter M. Tzelios}
\author{Kyle J. M. Bishop\footnote{Address correspondence to kyle.bishop@columbia.edu}~}
\affil{Department of Chemical Engineering\\Columbia University, New York, NY 10027}

\begin{document}

\maketitle

\begin{abstract}
Life on earth is distinguished by long-lived correlations in time.  The patterns of material organization that characterize the operation of living organisms today are contingent on events that took place billions of years ago. We and other living organisms are patterns that persist. This contingency is a necessary component of Darwinian evolution: patterns in the present inherit some of their features from those in the past. Despite its central role in biology, heritable information is difficult to recognize in prebiotic systems described in the language of chemistry or physics. Here, we consider one such description based on continuous time Markov processes and investigate the persistence of heritable information within large sets of dynamical systems. While the microscopic state of each system fluctuates incessantly, there exist few systems that relax slowly to their stationary distribution over much longer times. These systems, selected for their persistence, are further distinguished by their mesoscopic organization, which allows for accurate course grained descriptions of their dynamics at long times. The slow relaxation of these stable patterns is made possible by dissipative currents fueled by thermodynamic gradients present in the surrounding reservoirs. We show how the rate of entropy production within a system sets an upper bound on the lifetime of its persistent patterns.  The observation that dissipation enables persistence suggests that other features of living systems such as homeostasis and autocatalysis may emerge as consequences of a more general principle based on heritable information.  We also consider the probability of finding persistence within large sets of dynamical systems. We show that physical constraints based on continuity and locality can strongly influence the probability of persistence and its dependence on system size. Finally, we describe how heritable information can be quantified in practice using universal compression algorithms. We demonstrate this approach on an experimental system of magnetically-driven, colloidal rollers and discuss the application of these methods to origins of life research.
\end{abstract}

%%%%%%%%%%%%%%%%%%%%%%%%%%%%%%%%%%%%%%%%%%%%%%%%%%%%%%%%%%%%%%%
%%%%%%%%%%%%%%%%%%%%%%%%%%%%%%%%%%%%%%%%%%%%%%%%%%%%%%%%%%%%%%%
%%%%%%%%%%%%%%%%%%%%%%%%%%%%%%%%%%%%%%%%%%%%%%%%%%%%%%%%%%%%%%%
\clearpage

``Wherever matter is so poised, arranged, and adjusted, as to continue in perpetual motion, and yet preserve a constancy in the forms, its situation must, of necessity, have all the same appearance of art and contrivance which we observe at present.''

---David Hume, Dialogs Concerning Natural Religion, Part VIII

\section*{Introduction}

What kind of explanation can we expect to find for the origins of life\autocite{schrodinger1944life, dyson1999origins,eigen2013strange, pross2016life, dennett2017bacteria, davies2019demon, kauffman2019world} as we know it? We would appreciate a detailed \textit{historical narrative} addressing what, when, where, and how our most distant ancestors came to be. What were the material components of pre-life? When and where on Earth or elsewhere were they found? How did their transient fluctuations give way to a branching sequence of correlated events spanning billions of years? Sadly, even the most important features of this history may never be known. Recognizing these limitations, we might instead welcome \textit{synthetic demonstrations} whereby key features of living systems are reproduced from simple abiotic ingredients. Synthetic models of life should adapt their organization to harness energy from the surroundings and fuel dissipative functions.\autocite{england2015dissipative,te2018dissipative, Dou2019} These models should maintain local conditions favorable to their operation to provide autonomy from an ever-changing environment.\autocite{he2012synthetic,lerch2020homeostasis} As even the most reliable machines deteriorate, life-like systems should make copies of themselves to preserve dynamic patterns across generations.\autocite{colomb2015exponential, he2017exponential} Such synthetic demonstrations of dissipative adaptation, homeostasis, and autocatalysis promise to provide useful insights into the operation of early life-like systems. However, the \emph{a priori} emphasis on key features of life risks obscuring the historical processes by which these features emerged. To better understand and perhaps recreate that process, we require an appropriate \textit{conceptual framework} with which to approach the physical history of life’s origins.

\begin{figure}[t]
    \centering
    \includegraphics{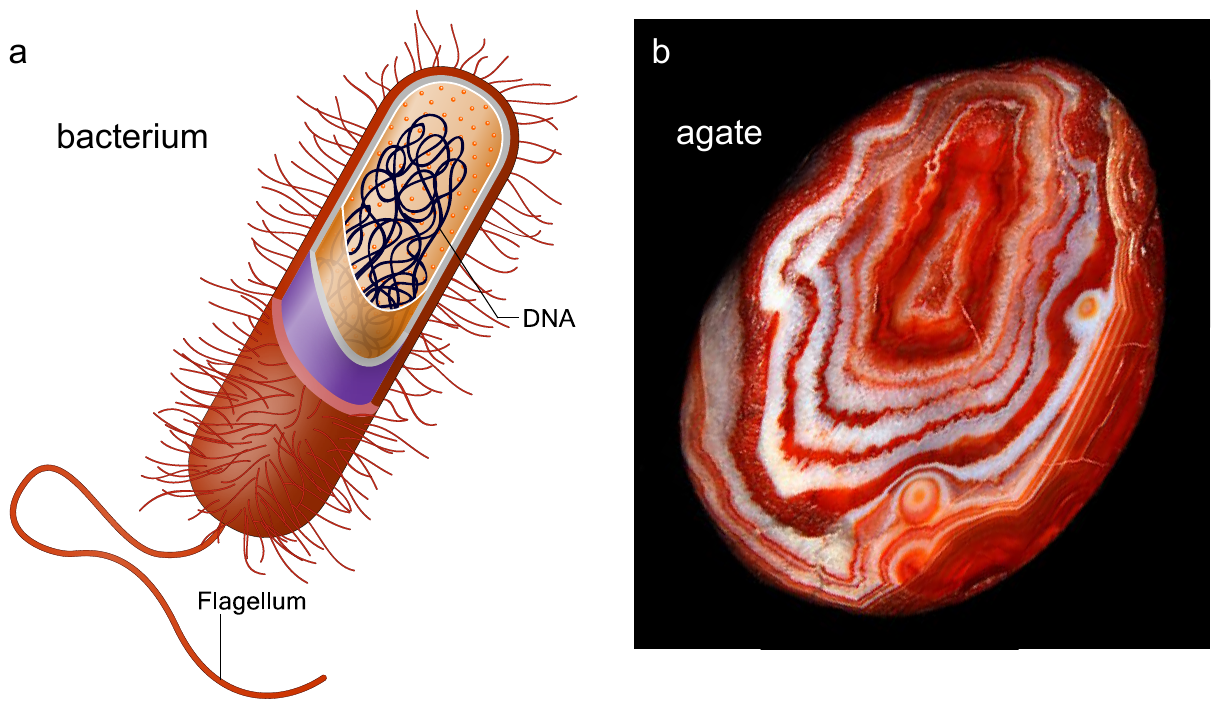}
    \caption{Examples of persistent patterns containing features correlated to past events \emph{ca.}~one billion years ago. (a) Schematic illustration of a bacterium highlighting the actuated appendage used to swim (flagellum) and its genetic material (DNA). Image adapted from work by \href{https://en.wikipedia.org/wiki/Prokaryote\#/media/File:Prokaryote_cell.svg}{Ali Zifan}. (b) Photograph of a Lake Superior agate. Image by \href{https://en.wikipedia.org/wiki/Lake_Superior_agate\#/media/File:Agat_Lake_Superior_-_Duluth,_Minnesota,_USA..jpg}{Lech Darski}.}
    \label{fig:intro}
\end{figure}

To illustrate the important role of such a framework, consider another origin---not of life---but of bacterial mobility (Figure \ref{fig:intro}a). How did bacteria “learn” to swim?  While we lack a full historical narrative of this transition, artifacts like the fossil record can provide clues. At one point in time, there were organisms recognizable as bacteria that lacked the biomolecular machinery needed to swim. Today, we observe bacteria that can swim. Moreover, we know how they do it---for example, how the flagellar motor converts proton gradients into rotary motion of helical filaments to propel bacteria through viscous fluids. Building on this knowledge, we’ve created synthetic demonstrations of primitive, micron-scale motors that swim like bacteria and unlike bacteria.\autocite{illien2017fuelled,bechinger2016active} However, these historical artifacts and synthetic demonstrations cannot explain the origin of bacterial mobility without the conceptual framework of Darwinian evolution by natural selection. Even without key details of this evolutionary transition, we are nevertheless confident that it can be explained by the differential survival and reproduction of bacteria exhibiting heritable variation.  We reason that swimming must be good for the survival of bacteria descendent from those that chanced upon this winning innovation. 

The central hypothesis of this Chapter is that the conceptual framework for understanding life’s origins is a suitable generalization of natural selection that describes the evolution of inanimate physical systems.\autocite{pross2011toward,Goldenfeld2011} The critical ingredient for such a framework is \emph{heritable information}.\autocite{schrodinger1944life} In biology, the replication of the genome provides the most important (though not the only) mechanism by which heritable information is passed from one generation to the next. An organism’s genome contains \emph{patterns}\autocite{dennett1991real} that persist over time---for example, the highly conserved genes shared by all living organisms. The persistence of these patterns over billions of years is remarkable considering the comparatively short lifetime of DNA---the physical medium in which genes are encoded. Genetic patterns are not just molecular structures but also dynamic processes that harness matter and energy in service of their persistence. Those patterns that survive the march of time capture our attention with ``the appearance of art and contrivance''; those that fade away and the many that never were do not.\autocite{dennett2017bacteria} Outside the context of biology, we will refer to this blind selection mechanism as the \emph{persistence filter} whereby time reveals those patterns that persist in a given environment.

The qualitative notion of patterns that persist can be quantified using dynamical models that describe the time evolution of physical systems. Here, we consider systems described by a finite number of discrete states—corresponding, for example, to the different possible configurations of system components. The system state changes in time due to stochastic transitions, which reflect uncertainties introduced by the fluctuating environment and by hidden variables neglected in the model dynamics. On sufficiently long times, these dynamics lead the system to explore all possible states with finite probability as determined by the system components and the surrounding environment. On shorter times, however, the state of the system is determined also by its history: knowledge of the present state (and of the system dynamics) teaches us something about the not-so-distant past.\autocite{keim2019memory} This heritable information can be quantified in terms of the mutual information between the current state of the system and its state at a previous time. All physical systems exhibit this simple form of heredity; however, only precious few exhibit life-like persistence.

Living systems are remarkable as their patterns persist on time scales far exceeding those of the microscopic processes from which they are made. By contrast, consider the persistent patterns of the Lake Superior agate---mineral formations created by geological events more than a billion years ago (Figure \ref{fig:intro}b). The banded deposits of iron-rich minerals are highly correlated across time for the simple reason that their atomic components are themselves unchanged. Meanwhile, some 98 percent of the atoms in the human body are replaced each year. The patterns that distinguish us as unique individuals---and those shared with other living organisms---exhibit a type of dynamical persistence exceedingly rare among physical systems. But just how improbable is life-like persistence within a prescribed class of physical systems? What are the mechanisms by which their physical histories are preserved? Under what conditions is dynamical persistence most likely to be found, and how will we recognize it?  This Chapter seeks answers---however tentative and incomplete---to these important questions.

We investigate the persistent patterns found within a set of dynamical systems based on continuous time Markov processes on a finite state space. To prohibit trivial forms of persistence---like that of the agate---we fix the lifetime of each state to unity. While the microscopic state of each system fluctuates incessantly, there exist few systems that relax slowly to their stationary distribution over much longer times. These systems, selected for their persistence, are further distinguished by their mesoscopic organization, which allows for accurate course grained descriptions of their dynamics at long times. The slow relaxation of these stable patterns is made possible by dissipative currents fueled by thermodynamic gradients present in the surrounding reservoirs. We show how the rate of entropy production within a system sets an upper bound on the lifetime of its persistent patterns. The observation that dissipation enables persistence suggests that other features of living systems such as homeostasis and autocatalysis may emerge as consequences of a more general principle based on heritable information. We also consider the probability of finding persistence within large sets of dynamical systems. We show that physical constraints based on continuity and locality can strongly influence the probability of persistence and its dependence on system size. Finally, we describe how heritable information can be quantified in practice using universal compression algorithms. We demonstrate this approach on an experimental system of magnetically-driven, colloidal rollers and discuss the application of these methods to origins of life research.

Overall, this Chapter advocates for a  physical perspective of life and its origins that places heredity at the center.  Importantly, the form of heredity we describe is not based on specific mechanisms such as self-replicating molecules but rather on the stable features of dynamical systems. This generalization of heritable information allows for the application of evolutionary concepts such as genotype and phenotype within even simple, non-biological systems. The cost of this generality, however, is the challenge of connecting abstract representations in terms of microstates and mesostates with the physical world of our experience. We therefore begin with a tutorial introduction to Markov state models covering the basic concepts needed for the present investigation. Building on insights from these abstract models, the final section aims to apply the ideas of this Chapter to an experimental model of driven colloids.  It is our hope that this work will inspire others to go searching for heritable information in other experimental systems from active matter\autocite{marchetti2013hydrodynamics} to systems chemistry\autocite{ashkenasy2017systems} and beyond.

%%%%%%%%%%%%%%%%%%%%%%%%%%%%%%%%%%%%%%%%%%%%%%%%%%%%%%%%%%%%%%%
%%%%%%%%%%%%%%%%%%%%%%%%%%%%%%%%%%%%%%%%%%%%%%%%%%%%%%%%%%%%%%%
%%%%%%%%%%%%%%%%%%%%%%%%%%%%%%%%%%%%%%%%%%%%%%%%%%%%%%%%%%%%%%%
\section*{Markov Processes}

To investigate the origins of dynamical persistence in physical systems, we consider a class of continuous time Markov processes on systems containing $N$ discrete states.\autocite{schnakenberg1976network}  The Markov property implies that the transition to state $i$ from state $j$ occurs with a rate $w_{ij}(t)$ independent of the system's history.\autocite{van1992stochastic}  This framework has been widely applied in describing the dynamics of physical systems across many scales including molecular dynamics,\autocite{Chodera2014,Husic2018} chemical kinetics,\autocite{Gillespie2007} and biomolecular machines.\autocite{Kolomeisky2007,Wagoner2016} The success of these models relies on their ability to average over fast microscopic processes and obtain coarse grained descriptions of slow mesoscopic processes of interest. For example, in describing the kinetics of chemical reactions, a coarse grained state may specify the numbers of different reacting species but not their microscopic positions and momenta nor those of the surrounding solvent molecules. Transitions from one state to the next are inherently stochastic due to fluctuations in the fast microscopic processes neglected under coarse graining. Additionally, these transitions are coupled to one or more external reservoirs that specify the temperature and chemical potential of the environment(s). The presence of multiple reservoirs out of equilibrium with one another allows for describing open thermodynamic systems characterized by currents of energy and species within and across the system.\autocite{Esposito2012,seifert2018stochastic} 
%There exists a complete thermodynamic formalism---so-called stochastic thermodynamics---that describes the fluctuating trajectories of Markov processes.

\begin{figure}[p]
    \centering
    \includegraphics{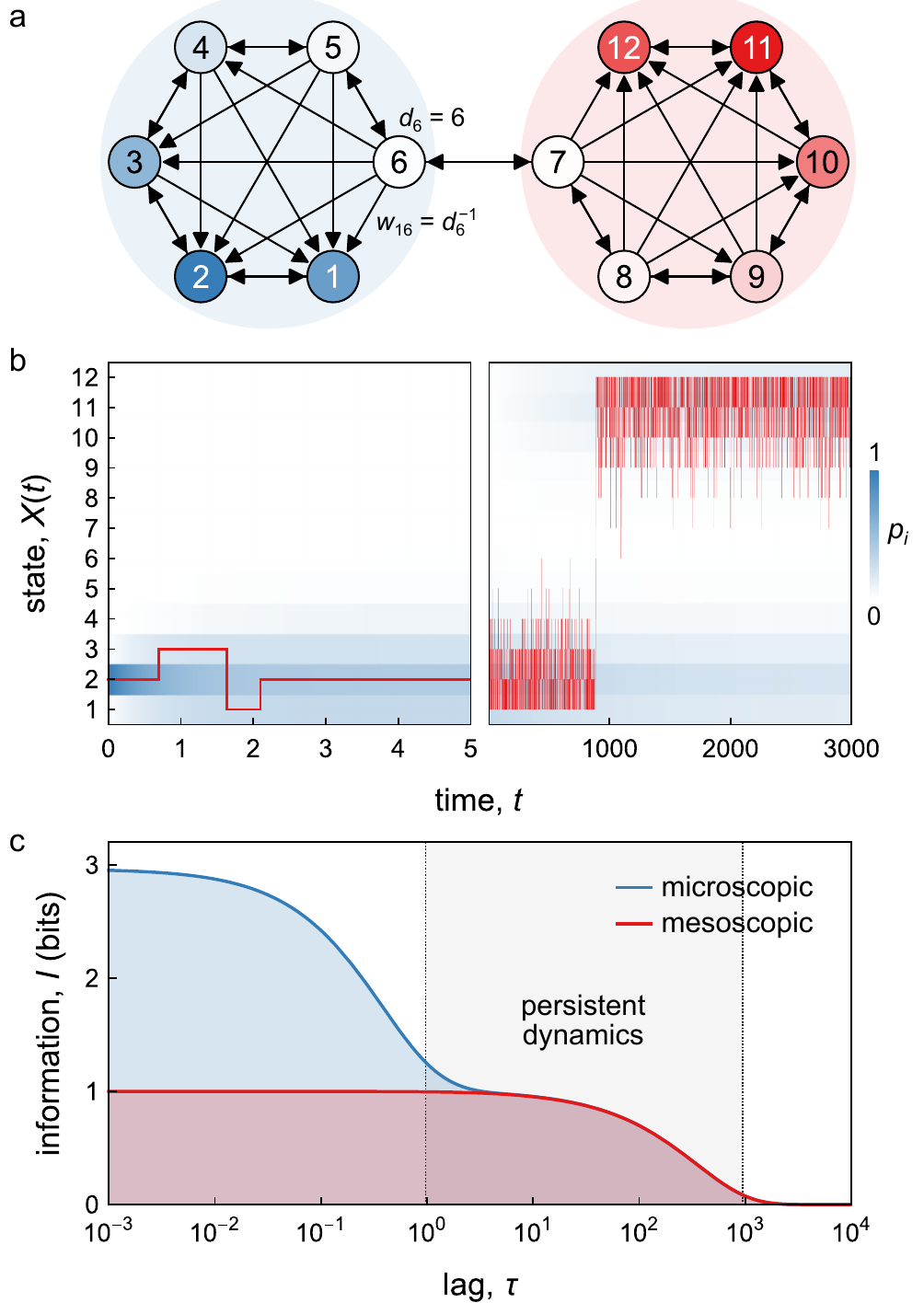}
    \caption{A 12-state system with persistent dynamics. (a) Graphical representation of the system. The numbered states are connected by directed edges representing stochastic transitions. An edge to state $i$ from state $j$ is characterized by a transition rate $w_{ij}=d_j^{-1}$, where $d_j$ is the out-degree of state $j$. Each state $j$ has an average lifetime of one by which other times are scaled---in other words, $\sum_i w_{ij}=1$.  At long times ($t\gg1$), the dynamics of the system is well approximated by a coarse grained description involving two mesostates (shaded circles). (b) Stochastic dynamics of the system. Starting from the initial condition $X(0)=2$, the plot shows one realization of the fluctuating state $X(t)$ as a function of time (red curve). The colormap shows the transient probability distribution $p_i(t)$ obtained by solving the master equation (\ref{eq:master}). (c) Mutual information between the past and the present.  On average, knowledge of the present state $X(t)$ provides $I(\tau)$ bits of information about the past state $X(t-\tau)$.  The plot shows how this information decays with increasing time lag $\tau$ from its maximum value, corresponding to the entropy of the stationary distribution $p_i(t\rightarrow\infty)=\pi_i$, to its asymptotic value of zero. The red curve shows the time-dependent information content associated with the coarse grained description.}
    \label{fig:Example}
\end{figure}

%%%%%%%%%%%%%%%%%%%%%%%%%%%%%%%%%%%%%%%%%%%%%%%%%%%%%%%%%%%%%%%
\paragraph{Stochastic Dynamics.}
The dynamics can be visualized as a random-walk on a graph comprised of $N$ states connected by directed edges representing stochastic transitions (Figure \ref{fig:Example}a).\autocite{schnakenberg1976network} Given that the system is in state $j$ at time $t$, it will jump to a different state $i$ with probability $w_{ij}(t)dt$ during a differential time interval $dt$. In general, the transition rates $w_{ij}$ depend on time---for example, due to changes in the surrounding environment; however, we limit our discussion to homogeneous processes with constant rates. Upon arriving in state $i$, the system waits for some time before jumping to another state $k$. The waiting time is exponentially distributed with a mean value of $(\sum_k w_{ki})^{-1}$ for state $i$. We focus on systems in which all states have a common lifetime by which other times are measured---that is, $\sum_k w_{ki}=1$ for all $i$. As shown below, the ability of such systems to preserve features of their past over long times ($t\gg1$) derives---not from local kinetic traps---but from mesoscopic organization and dissipative currents.

We consider a discrete set of possible dynamics parameterized by a directed adjacency matrix $A$. This binary matrix has elements $A_{ij}=1$ when there exists a transition to state $i$ from state $j$ and $A_{ij}=0$ otherwise. The associated transition rate is specified as $w_{ij} = A_{ij}/d_j$ where $d_j=\sum_i A_{ij}$ is the out degree of state $i$.  In other words, all transitions exiting state $i$ proceed at a common rate such that its mean lifetime is unity. Of the $2^{N(N-1)}$ possible adjacency matrices, we consider only those that encode strongly connected graphs, for which there exists a directed path from $i$ to $j$ and from $j$ to $i$ for every pair of states $(i,j)$. Despite the simplifying assumptions of equal rates and unit lifetimes, this space of possible dynamics is sufficiently large to include some of the complex dynamical behaviors expected in real systems. To facilitate a comprehensive exploration of these dynamics, we limit our initial investigation to the ca.~$10^{114}$ possible systems with less than or equal to $N=20$ states. 

%%%%%%%%%%%%%%%%%%%%%%%%%%%%%%%%%%%%%%%%%%%%%%%%%%%%%%%%%%%%%%%
\paragraph{Master Equation.}
The time-dependent probability distribution over the $N$ states is governed by the so-called master equation
\begin{equation}
    \frac{d p_i}{d t} = \sum_{j} W_{ij} p_j \label{eq:master}
\end{equation}
where $p_i(t)\equiv \Pr(X(t)=i)$ denotes the probability that the system is in state $i$ at time $t$, and $W_{ij} = w_{ij} - \delta_{ij} \sum_k w_{ki}$ are the elements of the transition rate matrix $W$. This linear equation describes the conservation of probability as it flows and spreads among the states of the system. 

To solve the master equation, we first decompose the rate matrix into its eigenvalues and eigenvectors as $W=-Q\Lambda Q^{-1}$, where $Q$ is the matrix of eigenvectors and $\Lambda$ is the diagonal matrix of eigenvalues with elements $\Lambda_{nn}=\lambda_n$. The transient probability distribution can then be expressed as a series of eigenmodes that decay independently as 
\begin{equation}
    p_i(t) = \sum_{j} T_{ij}(t) p_j(0) \quad \text{with} \quad T_{ij}(t) = \sum_n Q_{in} e^{-\lambda_n t} Q_{nj}^{-1} \label{eq:pi}
\end{equation}
where $p_j(0)$ is the initial distribution at time $t=0$ (Figure \ref{fig:Example}b). The matrix $T(t)$ describes a linear transformation by which the initial probability vector $\ve{p}(0)$ is squeezed, rotated, and sheared to produce the distribution $\ve{p}(t)$ at time $t$.  For a strongly connected graph with constant transition rates, exactly one of the eigenvalues is equal to zero (denoted $\lambda_1$). The associated eigenmode corresponds to the unique steady-state solution, $p_i(t\rightarrow\infty)= \pi_i = Q_{i1} / \sum_i Q_{i1}$.\autocite{schnakenberg1976network} The other eigenvalues have positive real parts, $\mathrm{Re}(\lambda_n)>0$, indicating that the stationary solution is stable. By convention, we number the eigenvalues in order of increasing real part such that $0=\lambda_1<\RE(\lambda_2)\leq \RE(\lambda_3)\leq \dots\leq\RE(\lambda_N)$.

%Conservation of probability requires that $\sum_i T_{ij}(t)=1$ for all columns $j$ and times $t$, which implies the following relationship for the $n^{\mathrm{th}}$ eigenvector, $\sum_i Q_{in} Q^{-1}_{nj}=\delta_{n1}$.

%%%%%%%%%%%%%%%%%%%%%%%%%%%%%%%%%%%%%%%%%%%%%%%%%%%%%%%%%%%%%%%
\paragraph{Dynamical Persistence.} 
Once the system relaxes to its steady-state distribution $\pi_i$, it retains no memory of its initial condition $p_i(0)$. We seek to identify persistent dynamics that retain features of their past for as long as possible (Figure \ref{fig:Example}c). This type of dynamical persistence can be quantified in terms of the mutual information $I(\tau)$ between the current state of the system $X(t)$ and its past state $X(t-\tau)$
\begin{equation}
    I(\tau) = H(X(t-\tau)) - H(X(t-\tau)\mid X(t)) \label{eq:I1}
\end{equation}
Here, $H(X(t-\tau))$ is the marginal entropy representing our uncertainty in the state of the system at time $t-\tau$. Upon observing the system's state at time $t$, we expect to learn something about its past thereby reducing our uncertainty. The conditional entropy $H(X(t-\tau)\mid X(t))$ represents the expected uncertainty about the system's past state given knowledge of its present state. Using equation (\ref{eq:pi}) for the distribution $p_i(t)$, the mutual information $I(\tau)$ can be written more explicitly as 
\begin{equation}
    I(\tau) = -\sum_i \pi_i \log_2 \pi_i +  \sum_{i,j} T_{ij}(\tau) \pi_j \log_2(T_{ij}(\tau)) \label{eq:I2}
\end{equation}
where the two terms in equation (\ref{eq:I2}) correspond to the entropies in equation (\ref{eq:I1}).  The use of binary logarithms implies that mutual information is measured in bits. 

Figure \ref{fig:Example}c shows how the amount of heritable information $I(\tau)$ decays with increasing lag time $\tau$ for a particular system with persistent dynamics.  For short lags ($\tau\ll 1$), knowledge of the present unambiguously determines the recent past such that $H(X(t-\tau)\mid X(t))\approx 0$; the mutual information equals the entropy of the steady-state distribution. For long lags, when even the slowest eigenmode has relaxed ($\RE(\lambda_2)\tau\gg1$), knowledge of the present provides no information about the past, and the mutual information approaches zero.  At intermediate times, $I(\tau)$ decreases monotonically with $\tau$ as the different eigenmodes relax in succession from fastest to slowest. 

In light of the relationship between mutual information and lag time, we can approach the question of persistence from two complementary perspectives. In one, persistent systems are those that preserve large amounts of information about their past for a specified lag time, $I(\tau)\geq 1$. Alternatively, we can invert this relationship and define persistent systems as those that preserve a specified amount of information for long times, $\tau(I)\gg1$.  For systems of given size $N$, there exists an inherent trade-off between the amount of heritable information $I$ and the time $\tau$ for which it is preserved.  We make use of both perspectives on persistence in exploring this trade-off for the set of dynamical systems introduced above. 

%%%%%%%%%%%%%%%%%%%%%%%%%%%%%%%%%%%%%%%%%%%%%%%%%%%%%%%%%%%%%%%
\paragraph{Coarse Graining.}
Systems with persistent dynamics preserve information about their past for long times, but it remains unclear how this information is encoded.  What features of the past persist to the present? To answer this question, we first note that systems with dynamical persistence are characterized by ``slow'' eigenmodes that relax more slowly than the remaining ``fast'' modes---often by orders of magnitude. At intermediate time scales, the fast modes are fully relaxed while the slow modes remain largely unchanged and determined by the system's history.  Importantly, the fast modes act to redistribute probability \emph{within} specific clusters of states---so-called mesostates---while the slow modes describe transitions \emph{between} mesostates.  Consequently, the separation of time scales implies the existence of coarse grained descriptions that characterize the slow dynamics of the system among the different mesostates. When the fast and slow times scales are strongly separated, transitions between the different mesostates are well approximated by Markov processes like those describing the original microscopic dynamics. 

Figure \ref{fig:Example} shows a system with $N=12$ states organized to form $M=2$ mesostates. While the average eigenvalue is equal to one, there exist one slow mode that relaxes more slowly with rate $\lambda_2\approx0.001$.  This mode describes the relaxation of the system between the two mesostates. On the slow time scale, the system is accurately described by a coarse grained description based on mesostate probabilities, $P_k(t)=\sum_i \Pi_{ki} p_i(t)$, where $P_k(t)\equiv\Pr(Y(t)=k)$ denotes the probability that the system is in mesostate $k$ at time $t$.  The conditional probability, $\Pi_{ki}\equiv\Pr(Y(t)=k\mid X(t)=i)$, serves to define the mesostate by associating each state $i$ to a mesostate $k$. To maximize the information content of this lossy description, we assign $\Pi_{ki}=1$ when state $i$ is in mesostate $k$ and $\Pi_{ki}=0$ otherwise (see Methods). In this way, the mutual information $I(\tau)$ computed from the mesostate probabilities accurately captures the persistent features of the system's dynamics (Figure  \ref{fig:Example}c).

%%%%%%%%%%%%%%%%%%%%%%%%%%%%%%%%%%%%%%%%%%%%%%%%%%%%%%%%%%%%%%%
\paragraph{Entropy Production.} 
The dynamical systems we investigate are, in general, out-of-equilibrium.  Consequently, they do not satisfy the condition of detailed balance, which requires the equality of forward and reverse rates at equilibrium, $w_{ij}\pi_j=w_{ji}\pi_i$. Instead, the steady-state distribution $\pi_i$ is characterized by dissipative currents that flow through the network of states due to thermodynamic forces imposed by the contacting reservoirs.  The rate of entropy production associated with these processes can be evaluated as  
\begin{equation}
    \sigma = \sum_{i,j} (w_{ij} \pi_j - w_{ji} \pi_i) \ln \left(\frac{w_{ij} \pi_j}{w_{ji} \pi_i}\right) \geq 0
\end{equation}
where equality holds at equilibrium. As we will show, the rate of entropy production sets bounds on persistence within our set of possible dynamics. In short, dissipation enables persistence. To explore this relationship, it is necessary to modify the dynamics to ensure that each nonzero transition rate, $w_{ij}\neq0$, is accompanied by another for the reverse process, $w_{ji}\neq0$.

We introduce a parameter $\gamma \in [0,1)$ that controls the extent of disequilibrium and allows for moving continuously between equilibrium and nonequilibrium dynamics.  Under conditions of maximum disequilibrium ($\gamma\rightarrow1$), transition rates are related to the adjacency matrix as $w^{\text{neq}}_{ij} = A_{ij}/d_j$ as described above. In the opposite limit ($\gamma=0$), the equilibrium transition rates are defined as $w^{\text{eq}}_{ij}=A^{\text{eq}}_{ij}/d^{\text{eq}}_j$ where $A^{\text{eq}}_{ij}$ is the symmetric adjacency matrix formed by the logical disjunction of the directed adjacency matrix and its inverse---that is, $A^{\text{eq}}_{ij}=A_{ij}\lor A_{ji}$.  For intermediate values of $\gamma$, the transition rates are equal to a weighted average of these limiting scenarios
\begin{equation}
    w_{ij} = w^{\text{eq}}_{ij} + \gamma (w^{\text{neq}}_{ij} - w^{\text{neq}}_{ij}) \label{eq:wij}
\end{equation}
When $\gamma=0$, the system satisfies detailed balance. The equilibrium distribution is  $\pi^\text{eq}_j\propto d^{\text{eq}}_j$, where $d_j^{\text{eq}}=\sum_i A^{\text{eq}}_{ij}$ is the degree of state $j$ within the equilibrium graph.  In the limit as $\gamma\rightarrow 1$, transitions between states become irreversible; the rate of entropy production diverges even as the dynamics remains well-behaved.  Unless otherwise noted, the results below correspond to systems under conditions of maximum disequilibrium, $\gamma\rightarrow1$.

%%%%%%%%%%%%%%%%%%%%%%%%%%%%%%%%%%%%%%%%%%%%%%%%%%%%%%%%%%%%%%%
%%%%%%%%%%%%%%%%%%%%%%%%%%%%%%%%%%%%%%%%%%%%%%%%%%%%%%%%%%%%%%%
%%%%%%%%%%%%%%%%%%%%%%%%%%%%%%%%%%%%%%%%%%%%%%%%%%%%%%%%%%%%%%%
\section*{Results}

%%%%%%%%%%%%%%%%%%%%%%%%%%%%%%%%%%%%%%%%%%%%%%%%%%%%%%%%%%%%%%%
%%%%%%%%%%%%%%%%%%%%%%%%%%%%%%%%%%%%%%%%%%%%%%%%%%%%%%%%%%%%%%%
\subsection*{The Persistence Filter} 

We consider a large number of systems with different rate matrices $W$, each initialized in a particular state drawn from the stationary distribution $\ve{\pi}$. These systems evolve independently of one another for a specified time $\tau$, longer than the microscopic time scale associated with each state (here, $\tau\gg1$). Upon observing the current states of the different systems, what do we learn about their respective histories?  For the vast majority, we learn nothing. The mutual information $I(\tau)$ between the past and the present has relaxed to zero. The state of such systems is fully explained by their components, their environment, and the whim of chance.  However, there exist few exceptional systems---distinguished by their dynamical persistence---for which the present state provides a window into the past. These systems are further characterized by their mesoscopic organization, which reveals ``clever'' strategies that enable their persistence.  This purposeful organization is designed by a blind selection mechanism---here, termed the \emph{persistence filter}---whereby time erases all but the most persistent patterns of dynamical organization. 

\begin{figure}[p]
    \centering
    \includegraphics{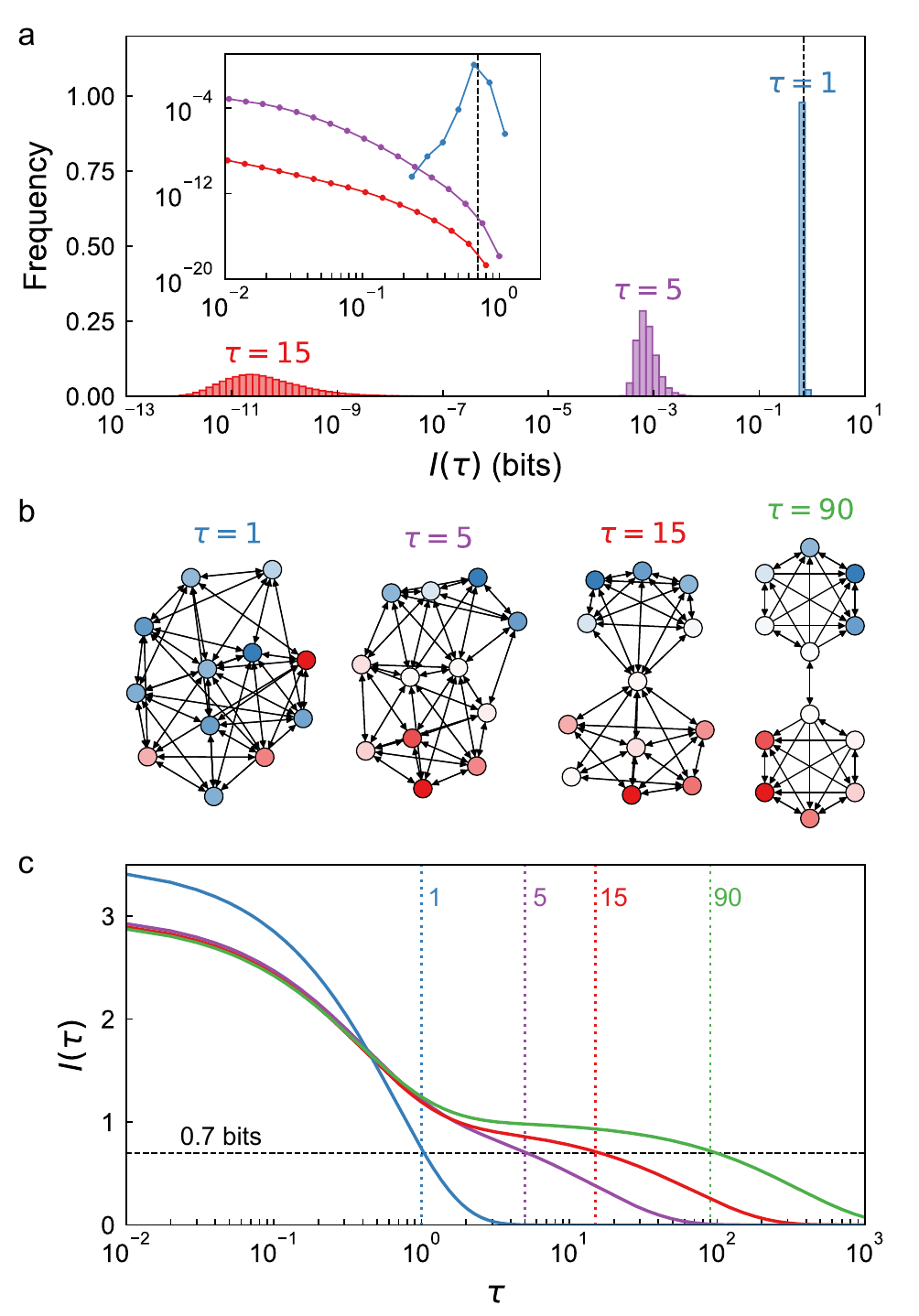}
    \caption{The persistence filter for systems with $N=12$ states. 
    (a) Distribution of mutual information $I(\tau)$ across all possible systems at different lag times $\tau$. Bins of the histograms are uniformly spaced on a logarithmic scale; the area under each curve is normalized to one. The inset shows the tail of the distribution corresponding to those systems with persistent dynamics---defined here as $I(\tau)\geq 0.7$ bits. 
    (b) Four systems sampled at random from those satisfying the constraint $I(\tau)\geq I_{\min}$ for $I_{\min}=0.7$ bits for different $\tau$. At long times, the sampled networks exhibit mesoscopic organization that enable their dynamical persistence.  
    (c) Mutual information for those systems in (b) as a function of lag time $\tau$. }
    \label{fig:PersistenceFilter}
\end{figure}

To demonstrate the persistence filter in action, we sample the space of possible dynamics for $N=12$ states to compute the distribution of information $I(\tau)$ for a specified lag time $\tau$ (Figure \ref{fig:PersistenceFilter}a). To efficiently sample rare systems with comparatively large $I(\tau)$, we use the Wang-Landau algorithm\autocite{Wang2001} originally developed for computing the density of states.  Initially at $\tau=0$, the amount of information averaged over the possible dynamics is $\langle I(0)\rangle = 3.5$ bits, which corresponds to the average entropy of the respective stationary distributions. As time progresses, the average information decays exponentially as the systems relax. The fraction of systems that encode one or more bits about their past also decreases exponentially.  By $\tau=15$, only one in ca.~$10^{20}$ systems is sufficiently persistent to preserve 0.7 bits of information.

Figure \ref{fig:PersistenceFilter}b shows four 12-state systems with persistent dynamics sampled from the tail of the information distribution at different lag times $\tau$. These systems are selected to preserve $0.7$ or more bits of information about their past state---that is, $I(\tau)\geq0.7$. As time progresses,  systems satisfying this constraint become increasingly rare within the set of ca.~$10^{39}$ possibilities. These exceptional systems are further distinguished by increasing levels of mesoscopic organization as evidenced by the modular structure of their network representations. This organization is a consequence of the slowly decaying eigenmodes selected by the persistence filter. By looking for information about the past, we are led to discover highly organized systems designed for the purpose of preserving that information. 

The persistence filter is analogous to Darwinian evolution by natural selection, in which the ``survival of the fittest'' is replaced by the less memorable aphorism ``persistence of the most persistent''.  Those patterns of dynamical organization determined by their history---not by system components and their environment alone---must persist in time if we are ever to recognize them. Moreover, the longer these patterns persist in spite of fast microscopic dynamics, the more organized the system must be to preserve them. Here, by ``organized'' we mean ``improbable''; only an exceptional few systems armed with clever strategies can effectively preserve heritable information. Below, we discuss some of these strategies and their respective frequencies within our set of possible dynamics.

\subsection*{Mechanisms of Persistence}

To better understand the mechanisms of dynamical persistence, it is helpful to consider the limiting cases---namely, those systems with $M$ modes that decay as slowly as possible. Specifically, we identify systems that minimize the $M^{\mathrm{th}}$ eigenvalue $\lambda_M$ of their respective rate matrices $W$. As detailed in the Methods, we approach this combinatorial optimization problem using Markov Chain Monte Carlo (MCMC) sampling with simulated annealing to explore the space of possible dynamics in search of those with the smallest $\lambda_M$.\autocite{kirkpatrick1983optimization}  For the dynamics considered here, the average eigenvalue is constant and equal to one. Thus, by minimizing $\lambda_{M}$, we identify systems with strong time-scale separation,  $\lambda_{M}\ll \lambda_{M+1}$, that allow for an accurate coarse grained description in terms of $M$ mesostates. 

The network in Figure \ref{fig:Example}a shows the optimal $12$-state system that minimizes the second eigenvalue $\lambda_2$. The system is divided into two mesostates, which are organized to prohibit transitions from one to the other (Figure \ref{fig:Example}a). Transitions between mesostates occur at a rate $\lambda_2=0.0011$, which is orders of magnitude slower than transitions within each mesostate (e.g., $\lambda_3=0.65$). Inspection of the network organization reveals the mechanism underlying this separation of time-scales. Within the blue mesostate, there exists a single linear path from state 1 to state 6, which is the sole gateway to the red mesostate. At each step along this path, competing transitions lead back to earlier states.  Only through a series of increasingly improbable events can the system escape these sisyphean cycles and transit to the opposite mesostate.

\begin{figure}[t]
    \centering
    \includegraphics{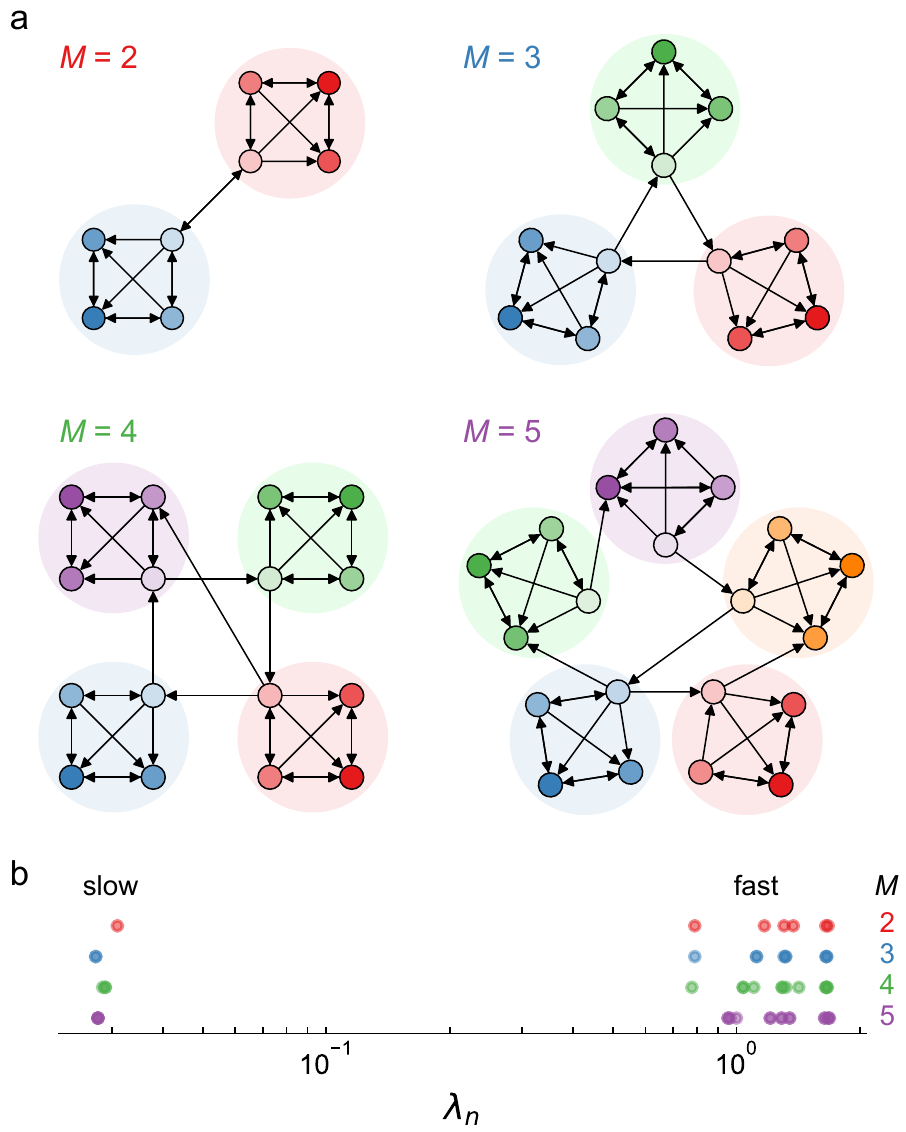}
    \caption{(a) Networks with $N=4 M$ states that minimize the eigenvalue $\lambda_M$ for $M=2,3,4,5$. Different colors are used to distinguish the $M$ mesostates; darker (lighter) shades denote higher (lower) probability at steady-state. (b) Eigenvalues for the networks in (a) highlighting time-scale separation between the slow modes and fast modes.}
    \label{fig:lambdaN}
\end{figure}

In general, minimizing the $M^{\mathrm{th}}$ eigenvalue $\lambda_{M}$ leads to the discovery of persistent systems with $M$ mesostates, which preserve ca.~$\log_2 M$ bits of information for as long as possible.  Figure \ref{fig:lambdaN} shows the result of this analysis for networks with $N=4M$ states for $M=2,3,4,5$.  We observe that the optimal networks are organized into $M$ mesostates of equal size (here, four states). As above, these clusters of states are organized internally to maximize their persistence and prohibit transitions to neighboring mesostates. As a result, the eigenvalues of the $M$ slow modes, $\lambda_n$ for $n=2,\dots,M$, have similar values comparable to the lifetime of each mesostate. By contrast, the remaining eigenvalues---those of the fast modes---are of order unity. 

For systems with three or more mesostates ($M\geq3$), transitions \emph{between} mesostates can also influence persistence; however, those \emph{within} mesostates are significantly more important. In Figure \ref{fig:lambdaN}, all mesostates within these maximally persistent systems share a common internal organization analogous to that discussed above for $M=2$ (cf.~Figure \ref{fig:Example}a). By contrast, the transitions between mesostates are loosely organized into sparse loops with no other discernible patterns.  Small changes in the network connectivity among mesostates serve to increase the relaxation rate $\lambda_M$ by ca.~10\%; such perturbations reduce persistence but only slightly. As the number of states grows, it becomes increasingly challenging to identify the global optimum in $\lambda_M$ among the ca.~$2^{N(N-1)}$ possibilities.  Fortunately, such technical details do not influence our central conclusion that dynamical persistence and network organization are inextricably linked.  Any form of selection for persistence will select also for mesoscopic organization similar to that shown in Figure \ref{fig:lambdaN}.  

Such coarse grained descriptions represent a type of \emph{pattern} whereby certain microscopic details are no longer required to describe or predict the system's mesoscopic behavior. At intermediate time scales ($\lambda_{M+1}^{-1}\ll t \ll \lambda_M^{-1}$), the system remains localized in one of many possible mesostates, which uniquely determines the quasi-steady distribution over its associated states.  This persistent pattern is analogous to the \emph{genotype} of a living organism, which persists from generation to generation with little change. Different genotypes specify different dynamical processes or \emph{phenotypes} in the service of maintaining their persistence. These processes include metabolism, motility, and reproduction among life's many remarkable innovations. Though comparatively primitive, the fluctuating currents within an occupied mesostate describe similar dissipative processes organized to foster its persistence.

%%%%%%%%%%%%%%%%%%%%%%%%%%%%%%%%%%%%%%%%%%%%%%%%%%%%%%%%%%%%%%%
\subsection*{Effects of Size $N$ and Disequilibrium $\gamma$}

\begin{figure}[t]
    \centering
    \includegraphics[width=\textwidth]{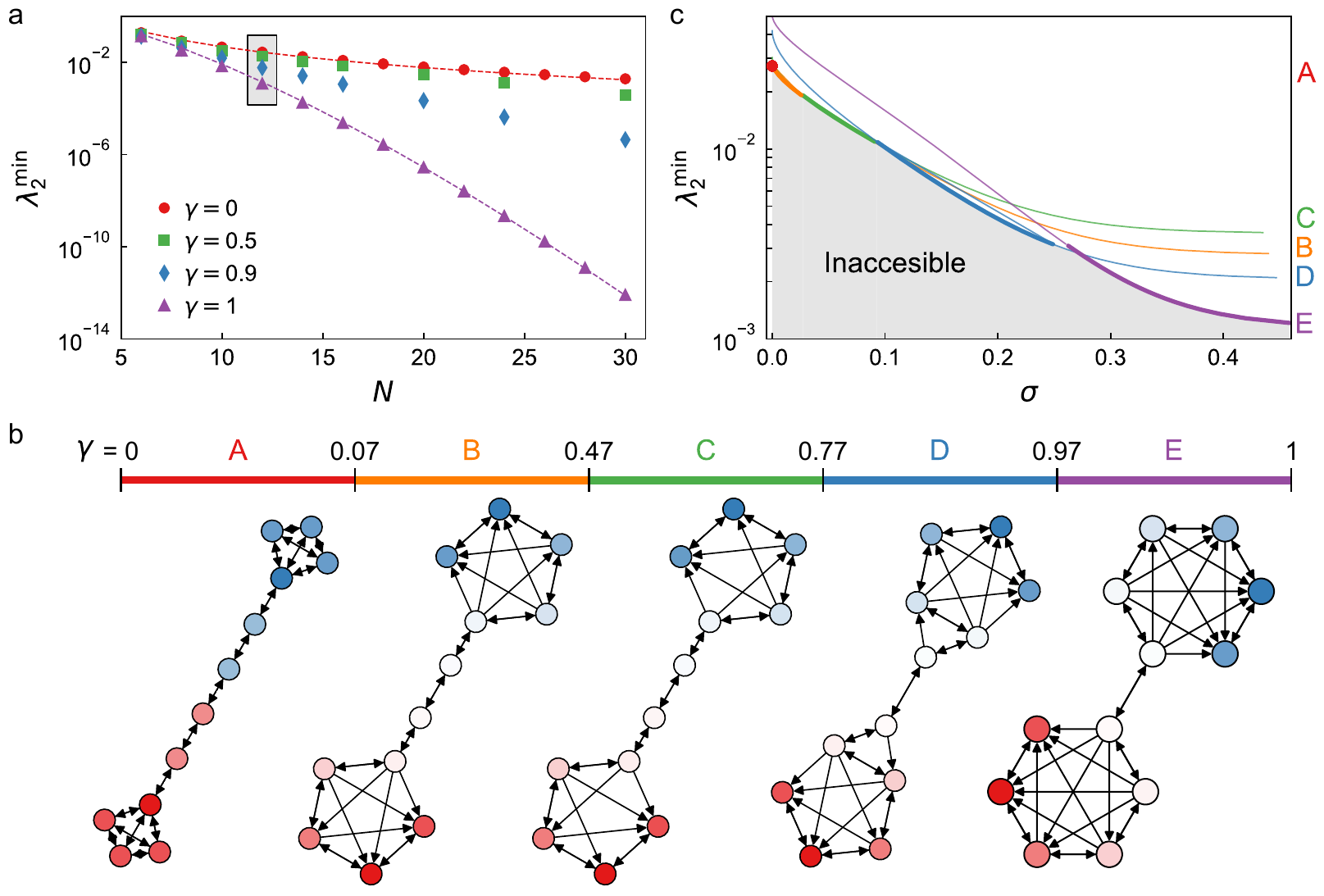}
    \caption{(a) Minimum eigenvalue $\lambda^{\min}_2$ as a function of network size $N$ for different degrees of disequilibrium $\gamma$. The solid markers correspond to specific networks identified by Monte Carlo optimization; the red dashed curve denotes the approximate relation for equilibrium networks, $\lambda_2^{\min}\approx 47 N^{-3}$; the purple dashed curve denotes the approximate relation for non-equilibrium networks with $\gamma\rightarrow1$, $\lambda_2^{\min}\approx1/\Gamma(N/2+1)$. 
    %\kbnote{[Change to $\lambda_2^{\min}$; italicize $N$; you might use different symbols for different $\gamma$ as well as different colors; add a box around the $N=12$ points.]} 
    (b) Network representations of  systems with $N=12$ states that minimize $\lambda_{2}^{\min}$ for different $\gamma$. The colored labels indicate the range of $\gamma$ for which each system is optimal; colors are matched to the boxed markers in (a). 
    %\kbnote{[Use a graphical representation for the ranges and label each network A, B, C, D, E as illustrated in my sketch below.]} 
    (c) Eigenvalues $\lambda_2$ and entropy production $\sigma$ for the systems in (b) as parameterized by the disequilibrium parameter $\gamma$. %\kbnote{[Can we try this one a bit different?  Plot each curve over the whole range of $\gamma$ as in my sketch below.  One key point is that there are regions of high persistence and low dissipation that are inaccessible.  ]}  
    }
    \label{fig:equilibrium}
\end{figure}

The lifetime of a mesostate optimized to persist increases faster than exponentially with the number states it contains. Figure \ref{fig:equilibrium}a illustrates this result for networks containing two mesostates that minimize $\lambda_2$.  The minimum eigenvalue $\lambda^{\min}_2$ characterizing the transition rate between mesostates decreases with the number of states as approximated by the relation, $\lambda^{\min}_2 \approx 1/\Gamma(N/2+1)$.  This result follows from the network architecture (Figure~\ref{fig:Example}a), which requires a specific series of transitions to escape the mesostate---namely, $i\rightarrow i+1$ with probability $1/i$ for $i=1,\dots,N/2$. With sufficient organization, as few as 48 states with ``microscopic'' lifetimes of order $10^{-42}$ s (the Planck time) can form a cluster with ``mesoscopic'' lifetimes of order $10^{17}$ s (the age of the universe).  Such persistence, however, has a thermodynamic cost that cannot be paid at equilibrium.

Equilibrium networks ($\gamma=0$) can also be organized to maximize persistence; however, the resulting relaxation rates scale algebraically (not exponentially) with the number of states. Figure \ref{fig:equilibrium}a shows that the minimum eigenvalue $\lambda_2^{\min}$ scales with network size $N$ as approximated by $\lambda_2^{\min}\approx 47 N^{-3}$ for equilibrium networks with $\gamma=0$ and $N\gg 1$ (see Appendix \ref{app:equil}).  These equilibrium networks are organized into two highly-connected clusters connected by a linear chain of states (Figure \ref{fig:equilibrium}b).  For large $N$, each cluster contains one third of the $N$ states (as opposed to one half for the non-equilibrium networks).  To achieve the same $10^{60}$ fold enhancement in persistence achieved by $N=48$ states for $\gamma\rightarrow 1$, an equilibrium network ($\gamma=0$) would require more than $N=10^{20}$ states.  We emphasize that these results are influenced by our starting assumption that all microstates have a common lifetime. Real systems---like that of the agate in Figure \ref{fig:intro}b---may contain low-energy states in which the system can be trapped for long times.

As $\gamma$ increases from zero to one, the system that maximizes persistence steps through a series of intermediate architectures between the limiting cases of equilibrium and maximum disequilibrium (Figure~\ref{fig:equilibrium}b). The different architectures correspond to different strategies for achieving persistence. The relative success of these strategies depends on the availability of thermodynamic resources needed to power dissipative currents. For nonequilibrium systems, these cyclic currents act to concentrate the occupancy probability deep within the interior of each mesostate, thereby preventing transitions between mesostates.  For networks of a given size, the rate of these transitions---as measured by the eigenvalue $\lambda_2$---decreases monotonically with the rate of entropy production $\sigma$ as parameterized by $\gamma$ (Figure~\ref{fig:equilibrium}c).  The observed relationship between these quantities suggests a thermodynamic cost\autocite{Dou2019} of persistence whereby the rate of entropy production bounds the maximum lifetime of each mesostate. Similar bounds on the precision of stochastic processes have been established\autocite{Barato2015,Gingrich2016} and provide a basis for thermodynamic inference\autocite{Horowitz2020} of unobserved properties in nonequilibrium systems.  

By searching for systems with persistent dynamics, we are led to discover nonequilibrium systems that make use of available resources to power their persistence.  However,  dissipative currents can also be used to erase information about the past and accelerate relaxation to the steady-state.  Thus, finite rates of entropy production are necessary but not sufficient to achieve persistence. This distinction is important in the context of life where features such as entropy production, compartmentalization, and replication among others are given privileged status in distinguishing life from non-life.  Here, we see that entropy production is a corollary of dynamical persistence; dissipative processes enable useful strategies for achieving persistence. Here, by ``useful'' we mean ``probable''; there are many more ways to preserve heritable information by harnessing available energy sources.  We hypothesize that other features of life such as compartmentalization and replication can be understood in similar fashion as consequences of persistence within appropriate classes of dynamical systems.

%%%%%%%%%%%%%%%%%%%%%%%%%%%%%%%%%%%%%%%%%%%%%%%%%%%%%%%%%%%%%%%
\subsection*{Probability of Persistence}

Over time, the persistence filter sifts through many dynamical systems to select those few that preserve information about their respective histories. The filter does not create these systems; instead, it focuses our attention on their unlikely existence among myriad alternatives with little or no history at all. To understand the existence of dynamical persistence---most notably that of life on Earth---it is instructive to ask how unlikely it is within a particular class of dynamical systems? Moreover, which types of systems are more likely to foster persistence? To address these questions, we search our space of possible dynamics to estimate the fraction of systems that encode one bit of information relaxing slower than a prescribed rate $\nu$. Specifically, we use the Wang-Landau algorithm\autocite{Wang2001} to compute the cumulative distribution $\Pr(\lambda_2 < \nu)$ for the eigenvalue $\lambda_2$. For now, each of the ca.~$2^{N(N-1)}$ possible dynamics is assumed to be equally likely. As discussed below, this prior distribution over the possible dynamics has a significant impact on the probability of finding persistent systems.

\begin{figure}[t]
    \centering
    \includegraphics{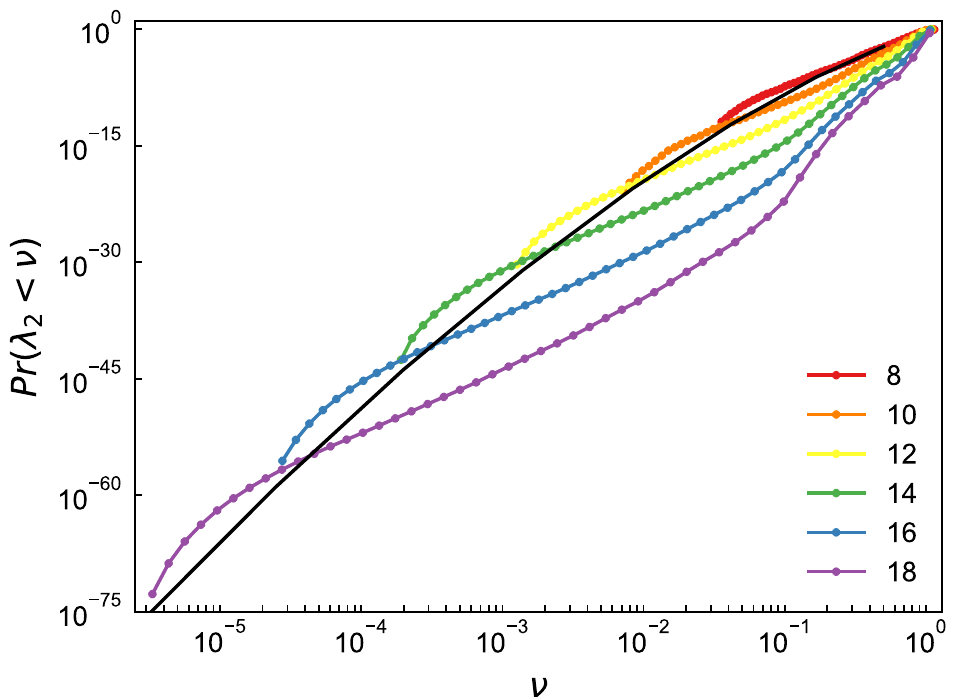}
    \caption{Probability of persistence.  The plot shows the cumulative probability $\Pr(\lambda_2<\nu)$ that the eigenvalue $\lambda_2$ is slower than a prescribed rate $\nu$. Colored markers correspond to results of the Wang-Landau calculations for systems of different sizes $N$. The black curve shows the approximate relation for systems with the maximal persistence---namely, $\lambda_2=\lambda_2^{\min}\approx1/\Gamma(N/2+1)$ and $\Pr(\lambda_2=\lambda_2^{\min})\approx N!/2^{N(N-1)}$.} 
    \label{fig:WL_scale}
\end{figure}

Figure \ref{fig:WL_scale}a shows the monotonic relationship between probability and persistence: systems with smaller eigenvalues $\lambda_2$ are increasingly hard to find. For a given size $N$, the eigenvalue $\lambda_2$ is bounded by its minimum value shown in Figure \ref{fig:equilibrium}a.  These maximally persistent systems correspond to $N!$ degenerate networks that differ only in the labeling of their states.  The probability of selecting one such system at random can therefore be approximated as $\Pr(\lambda_2=\lambda_2^{\min})\approx N!/2^{N(N-1)}$. This estimate represents a lower bound as it includes all possible adjacency matrices---even those that are not strongly connected. The black curve in Figure \ref{fig:WL_scale} shows that these approximate results for maximally persistent systems are representative of the general trends.  

For a given relaxation rate $\nu$, the probability of finding a persistent system decreases with increasing system size $N$ (Figure \ref{fig:WL_scale}).  The number of non-persistent systems grows faster with $N$ than does the number of persistent systems.  This observation suggests that persistence is most likely to be achieved using the smallest---that is, the simplest---system possible. The chances of finding dynamical persistence in macroscopic systems becomes vanishingly small, and yet, here we are. To reconcile this apparent contradiction, we must recognize that our prior distribution over the space of possible systems is a poor representation of physical reality. In particular, the assumption that all adjacency matrices are equally likely lacks the features of \emph{continuity} and \emph{locality} characteristic of physical dynamics. 

\begin{figure}[p]
    \centering
    \includegraphics{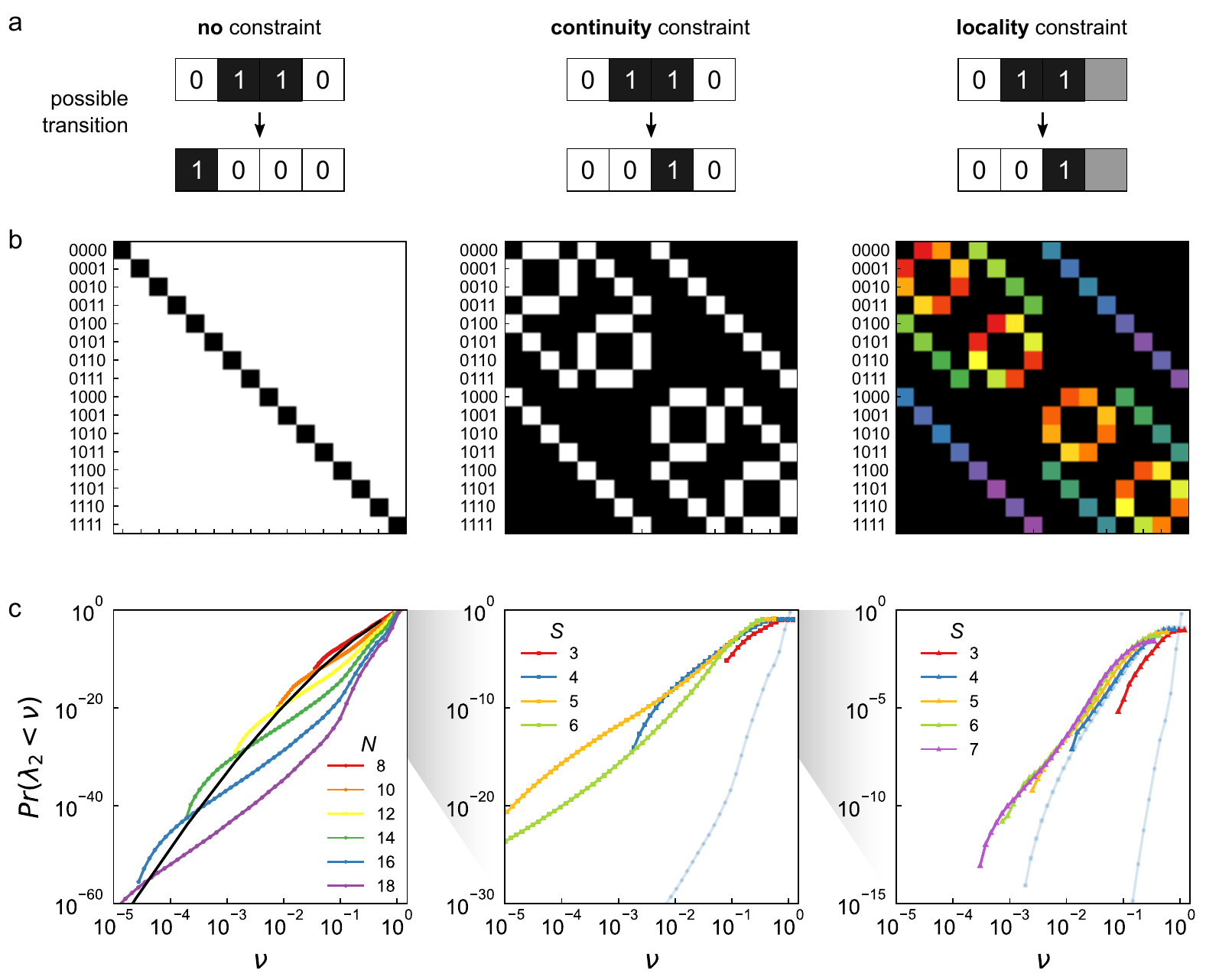}
    \caption{Constraints of continuity and locality alter the probability of persistence. (a) For $N=2^S$ states, we can associate each state with a sequence of $S$ binary variables or ``spins''. In the absence of constraints, transitions are allowed between any two states (left).  With the continuity constraint, only transitions that flip one spin are allowed (center).  Further addition of the locality constraint implies that the flipping of a spin depends only on its nearest neighbors (right). (b) Adjacency matrices $A$ illustrate the possible transitions allowed (white) for $N=16$ ($S=4$). For the locality constraint (right), possible transitions with the same color must take the same value in the adjacency matrix. (c)  Cumulative probability $\Pr(\lambda_2<\nu)$ that the eigenvalue $\lambda_2$ is slower than a prescribed rate $\nu$. The data from Figure \ref{fig:WL_scale} is reproduced (left) to facilitate comparison with the results of the constrained dynamics (center, right). Similarly, data for $N=16$ ($S=4$) is reproduced on successive plots to highlight the influence of the constraints.}
    \label{fig:constraints}
\end{figure}

\paragraph{Continuity Constraint.}
The principle of continuity implies that a microscopic transition to state $i$ from state $j$ is more likely to occur if those states are ``close'' to one another in phase space.  As a specific example, we consider a digital phase space of $S$ binary variables (or ``spins'') where each of the $N=2^S$ total states corresponds to a binary sequence of length $S$ (e.g., $01001\dots$). Microscopic transitions from one state to another are more likely to occur when those states differ by only a few spins. Here, we assume that each state can transition to $S$ possible neighboring states by flipping only one spin; transitions that involve the flipping of two or more spins are prohibited.  As a result, the number of possible dynamics (adjacency matrices) is reduced from ca.~$2^{N(N-1)}$ down to ca.~$2^{N \log_2 N}$ (Figure \ref{fig:constraints}b, center). Owing to the continuity constraint, large changes in phase space can occur only through sequences of many small changes.

Within the reduced space of possible dynamics, the probability of finding persistent systems increases by orders of magnitude (Figure \ref{fig:constraints}, center). For systems of $N=16$ states ($S=4$ spins), the fraction with two mesostates that relax slower than $\lambda_2<0.003$ increases from ca.~$10^{-34}$ to $10^{-12}$ on applying the continuity constraint. This increase is due to the significant reduction in the total number of systems, most of which are highly connected and non-persistent. The continuity constraint also excludes persistent systems but to lesser extent. For $N=16$, the minimum eigenvalue $\lambda_2^{\min}$ increases from $2.2\times10^{-5}$ to $1.5\times10^{-3}$ when excluding those systems that violate the continuity principle. As with any optimization, the addition of constraints necessarily increases the objective function at the minimum. More importantly, these results show how physical constraints on the space of possible dynamics can serve to make persistence more probable.

\paragraph{Locality Constraint.}
The principle of locality further implies that transitions between neighboring states in phase space are largely independent of system features that are ``far'' away in physical space. Continuing our example, the rate of flipping a particular spin may depend on neighboring spins but not on the detailed configurations of those far away in physical space. The exclusion of systems with non-local dynamics further constrains the space of possible dynamics and can therefore impact conclusions regarding the probability of persistence. Here, we arrange the $S$ spins within a linear array and limit the range of interactions to nearest neighbors (Figure \ref{fig:constraints}, right). Thus, the rate of flipping each spin depends only on the configuration of two neighboring spins and is independent of all other spins.  This constraint further reduces the number of possible dynamics from ca.~$2^{N \log_2 N}$ down to ca.~$2^{8 \log_2 N}$ for nearest neighbor interactions in 1D.

The rightmost plot of Figure \ref{fig:constraints}c shows the probability of persistence for systems satisfying the continuity and locality constraints described above. Owing to the significant reduction in the total number of systems, those satisfying these constraints require more states to achieve high levels of persistence.  On applying the locality constraint, the probability of persistence increases as compared to systems satisfying the continuity constraint alone---but not by much.  More important, however, is the dependence on system size $N$. For systems satisfying both continuity and locality, the probability of persistence appears to increase with the number of states $N$. These results are rather anecdotal---applying only to numerical results of a particular model for small systems---but are nevertheless suggestive of an important hypothesis. For certain distributions over possible dynamics within physical systems, persistent dynamics---of one form or another---are all but guaranteed for sufficiently large systems.  For other distributions, however, persistence is all but impossible as $N$ grows to the astronomical values typical of macroscopic systems.  This hypothesis has implications for the search for abiotic systems that exhibit characteristics of life.

\begin{figure}[t]
    \centering
    \includegraphics{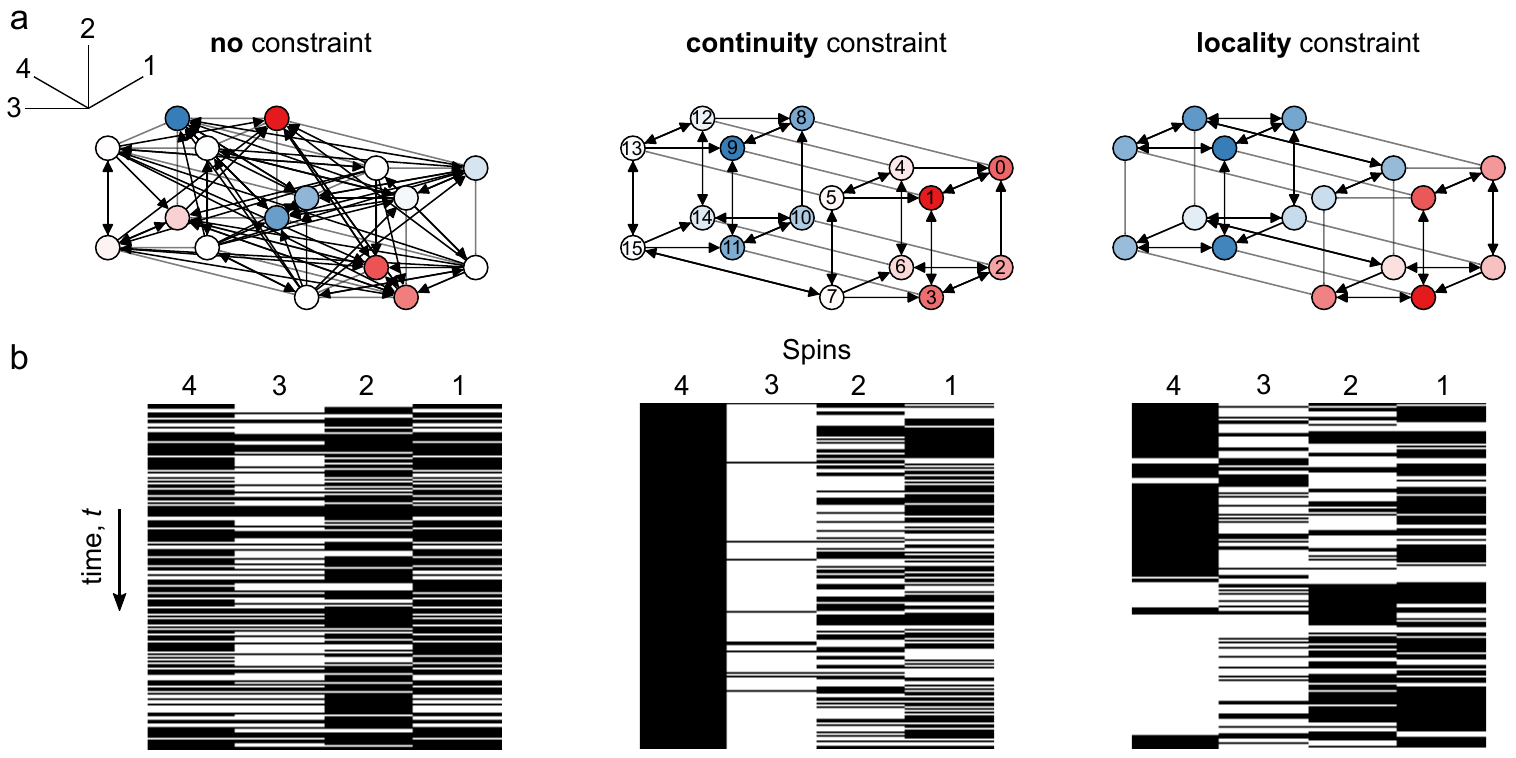}
    \caption{New strategies for persistence. (a) Network representations of systems with $S=4$ spins that minimize $\lambda_{2}$ for the physical constraints of Figure \ref{fig:constraints}. Each spin corresponds to a particular dimension of the 4D hypercube. (b) State-time images showing the system state as a function of time for the networks in (a). Each image is obtained from a single stochastic trajectory from $t=0$ to $t=500$ with a common initial condition.}
    \label{fig:new_strategy}
\end{figure}

\paragraph{New Strategies for Persistence.}
With the application of the continuity and locality constraints, the newly optimal systems adopt different strategies to maximize their persistence. Figure \ref{fig:new_strategy}a shows three such systems with 16 states and different constraints that minimize the second eigenvalue $\lambda_2$, thereby preserving one bit of information for as long as possible.  To facilitate comparison between these systems, the positions of the $2^4$ states are fixed at the corners of an $4$-dimensional hypercube; the networks differ only in their directed transitions. 

With no constraints, the maximally persistent system is analogous to that shown in Figure \ref{fig:Example}a; however, the network organization is obscured by the hypercube representation. There is no meaningful relationship between the $S=4$ spins and the $N=2^S$ states.  As a result, the organization of the the system into two mesostates is not at all apparent when visualized through the dynamics of the four spins (Figure \ref{fig:new_strategy}b, left), which fail to provide a useful description of the persistent pattern. This pattern is nevertheless present: the system remains localized in the blue mesostate throughout the duration of the simulation pictured here.  

By contrast, for systems that satisfy the continuity constraint, the identity of the mesostate is clearly encoded by a single variable in the phase space (Figure \ref{fig:new_strategy}a, center). In this way, the persistent dynamics of the system is immediately visible from the dynamics of the four spins (Figure \ref{fig:new_strategy}b, center).  The continuity constraint requires that transitions between states occur along one dimension of the hypercube at a time. Subject to this constraint, the strategy for achieving dynamical persistence is qualitatively similar to that of the unconstrained system.  Within each mesostate, there exists a single linear path to escape to the other mesostate (e.g., $0000\rightarrow0001\rightarrow0011\rightarrow0010\rightarrow0110\rightarrow0100\rightarrow0101\rightarrow0111\rightarrow1111$). At each step along this path, competing transitions lead back to earlier states, when permitted by the physical constraints. Further addition of the locality constraint results in similar networks that encode the mesostate within a single physical spin (Figure \ref{fig:new_strategy}b, right). In this way, the search for persistent dynamics leads to error correction mechanisms that preserve the configuration of one spin. More generally, this result suggests how physical constraints serve to concentrate mesoscopic information within few variables of a larger phase space.

%%%%%%%%%%%%%%%%%%%%%%%%%%%%%%%%%%%%%%%%%%%%%%%%%%%%%%%%%%%%%%%
%%%%%%%%%%%%%%%%%%%%%%%%%%%%%%%%%%%%%%%%%%%%%%%%%%%%%%%%%%%%%%%
\subsection*{Measuring Persistence in Practice}

For the idealized models described above, we quantify dynamical persistence using the mutual information between the system state at two different times. This approach requires knowledge of the system's microscopic dynamics in order to distinguish patterns associated with its history from those determined by its components and environment alone.  In practice, however, we don't know the underlying dynamics, and our measurements of the system state are incomplete. How then can we expect to recognize and quantify persistent dynamics in experimental models of life-like systems? Here, we described one approach based on computable information density (CID)\autocite{Martiniani2019} and apply it to an experimental model of driven colloidal particles.

\begin{figure}[t!]
    \centering
    \includegraphics{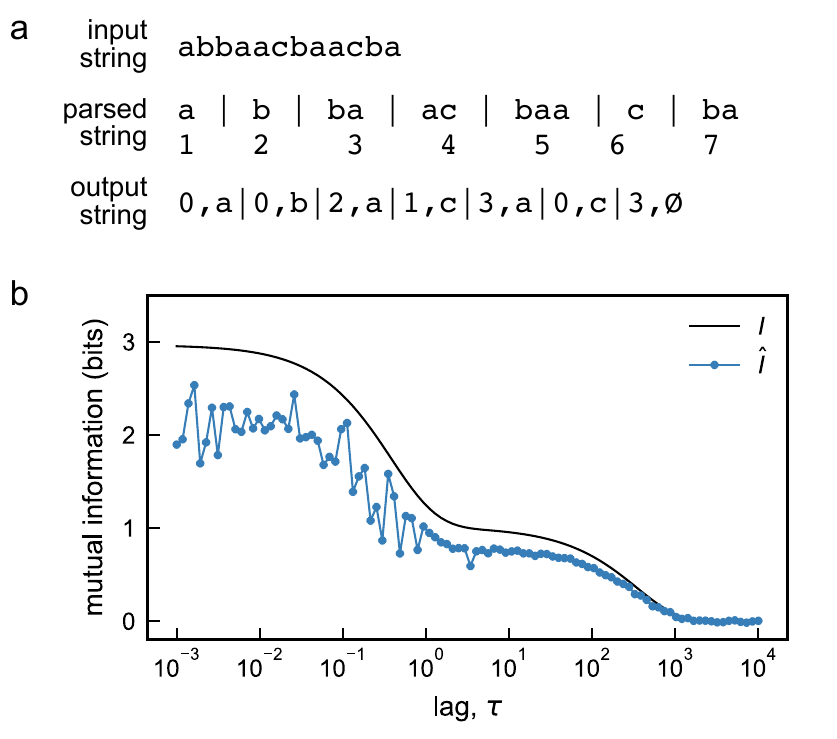}
    \caption{(a) Example of the tree-structured Lempel-Ziv compression algorithm (LZ78).\autocite{cover1999elements} A string of length $n=12$ with an alphabet of size $N=3$ is parsed into a sequence of $c(n)=7$ unique phrases. The compressed output contains the sequence of phrases, each specified by the index of their prefix phrase and their terminal letter. (b) The mutual information $I(\tau)$ for the 12-state system in Figure \ref{fig:equilibrium} (solid curve) is approximated by $\hat{I}(\tau)$ (markers) using sequences of $n=10^4$ states generated with a prescribed time step $\tau$.  All realizations of the stochastic process are initialized from state 1.}
    \label{fig:CID}
\end{figure}

\paragraph{Computable Information Density (CID).} The mutual information $I(\tau)$ between two system states separated by a time lag $\tau$ is the difference of two entropies: that of the stationary distribution $H(X(t))$ and that of the future state conditioned on the past $H(X(t)\mid X(t-\tau))$.  By observing the system state at regular time intervals $\tau$, we can estimate these quantities without explicit knowledge of the system dynamics.  The observed sequence, $x=\{X_1,X_2,\dots,X_n\}$, is analogous to a text string where each letter denotes a possible state of the system (Figure \ref{fig:CID}a). The desired entropies determine our ability to compress this string to form a more concise but nevertheless exact representation of the state sequence. Consequently, we can estimate the entropy of a string from the length of its binary representation after lossless compression.  The computable information density (CID) is defined as the ratio between the binary code length $\mathcal{L}(x)$ of the compressed string and the length $n$ of the original string
\begin{equation}
    \text{CID}(x)=\frac{\mathcal{L}(x)}{n} \label{eq:CID}
\end{equation}
This quantity has been used previously as an order parameter to describe phase transitions\autocite{Martiniani2019} and correlation lengths\autocite{Martiniani2020} in nonequilibrium systems. Here, we use the CID to measure time correlations and quantify  dynamical persistence in Markov processes.

We perform lossless compression of state sequences using the tree-structured Lempel–Ziv algorithm (LZ78)\autocite{Ziv1978,cover1999elements} as illustrated by the example in Figure \ref{fig:CID}a.  Given a string of length $n$ from a finite alphabet of size $N$, the algorithm parses the string into a sequence of ``phrases'' where the next phrase is the shortest phrase not seen in the past.  In the compressed representation, each phrase can be described by a pointer to the corresponding prefix phrase and by the terminal letter that completes the phrase. If there are $c(n)$ phrases in a particular string of length $n$, then the specification of the pointer requires at most $ \log_2 c(n) $ bits and the terminal letter at most $\log_2 N$ bits.  Therefore, the binary length of the compressed string is bounded as 
\begin{equation}
    \mathcal{L}(x) \leq c(n)(\log_2 c(n) + \log_2 N) \label{eq:bound}
\end{equation}
We use this bound to approximate the binary code length in evaluating the CID.

The LZ78 code---like LZ77 used in prior work\autocite{Martiniani2019,Martiniani2020}---is ``universal'' in that it compresses any sequence without \emph{a priori} knowledge of the process by which it was generated. Moreover, the code is ``optimal'' in that the $\mathrm{CID}(x)$ converges to the entropy rate for a  stationary ergodic process\autocite{cover1999elements}
\begin{equation}
    \lim_{n\rightarrow\infty}\mathrm{CID}(x) = \lim_{n\rightarrow\infty} \frac{1}{n} H(X_1,X_2,\dots,X_n) 
\end{equation}
For Markov processes in particular, the entropy rate is identical to the conditional entropy $H(X(t)\mid X(t-\tau))$ introduced above.\autocite{cover1999elements} In this way, the CID provides a convenient estimate for the conditional entropy, which is exact in the limit of long sequences. We note, however, that the difference between the expected CID and the entropy rate---the so-called redundancy---decays slowly as $(\log_2 n)^{-1}$.\autocite{Savari1997}

To estimate the marginal entropy $H(X(t))$, we shuffle the state sequence $x$ and evaluate the CID of the randomized sequence $x_{\sh}$.\autocite{cavagna2020vicsek} The process of shuffling acts to destroy time correlations between successive states.  For long times ($n \tau\gg \lambda_1^{-1}$), the shuffled sequence becomes indistinguishable from a set of $n$ independent samples drawn from the stationary distribution. With these preliminaries, the mutual information $I(\tau)$ can be estimated from state sequence $x$ with sample time $\tau$ as 
\begin{equation}
    I(\tau) \approx \hat{I}(\tau) = \langle \mathrm{CID(x_{\sh})} \rangle - \mathrm{CID(x)}  \label{eq:Icid}
\end{equation}
where $ \langle~\rangle$ denotes an average over the shuffled sequences. Figure \ref{fig:CID}b compares this computable approximation based on observed states to the exact result based on knowledge of the dynamics for the 12-state system of Figure \ref{fig:Example}.  Deviations between the two derive from the finite number of sampled states (here, $n=10^4$) and the finite sample times for short lags $\tau< (n \lambda_1)^{-1}\approx 0.1$.  

\begin{figure}[t]
    \centering
    \includegraphics{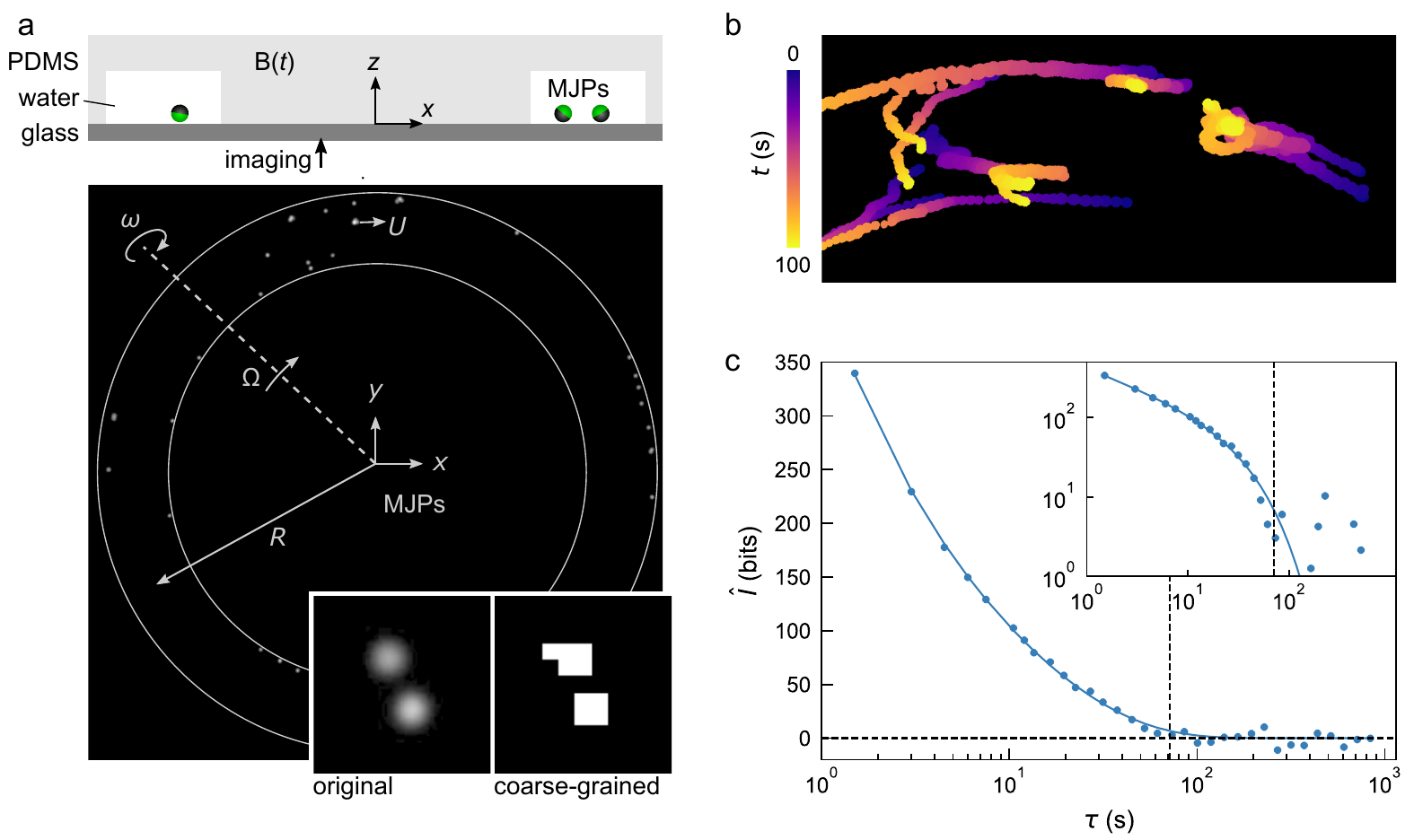}
    \caption{(a) Magnetic Janus particles (MJPs; 4 $\mu$m radius) are driven to roll around a circular track of radius $R = 400~\mu$m, width $100~\mu$m, and height $50~\mu$m by a time-varying magnetic field $\ve{B}(t)$. Particles and clusters thereof appear as bright spots on fluorescent microscopy images---here, annotated to show the circular track. The inset shows the downsampling and thresholding of the raw image to create the coarse grained representation used in our analysis.  (b) Time-lapse images show cluster dynamics within the moving front. Here, we adopt a rotating frame of reference in which the average position of the front is stationary.  (c) Mutual information estimate $\hat{I}(\tau)$ as a function of lag time $\tau$ for the system shown in (a). Here, the driving field is specified by equation (\ref{eq:B}) with magnitude $B_0=3$ mT, rolling frequency $\omega=62.8$ rad/s, and rotation frequency $\Omega=0.0140$ rad/s; the video is captured over 70 minutes with a frame rate of 0.67 fps. The markers denote the estimates computed from the videos; the solid curve is a least-squares fit of the form $C \tau^{-\alpha} e^{-\beta \tau}$ with $\alpha=0.49$ and $\beta=0.029$ s$^{-1}$. The inset shows the same data on a log-log plot.}
    \label{fig:rollers}
\end{figure}

\paragraph{Quantifying Persistence in Dynamic Assemblies of Colloidal Rollers.}
CID-based estimates of mutual information can be used to quantify dynamical persistence in experimental models of nonequilibrium systems.  Here, we consider one such model based on colloidal particles driven to roll along a solid surface by a time-varying magnetic field (Figure \ref{fig:rollers}a; see Methods for details).\autocite{driscoll2017unstable, dou2020programmable} The particles are confined within a circular track and driven to roll along circular orbits by fields of the form
\begin{equation}
    \ve{B}(t) = B_0\left(\sin(\Omega t)\sin(\omega t)\ve{e}_x + 
    \cos(\Omega t)\sin(\omega t) \ve{e}_y +
    \cos(\omega t) \ve{e}_z\right) \label{eq:B}
\end{equation}
where $B_0$ is the constant magnitude of the field, $\omega$ is the rolling frequency, and $\Omega$ is the comparatively slow frequency with which the rolling direction changes. This rotating field drives ferromagnetic spheres to roll across the surface with speed $U$ around circular orbits of radius $U/\Omega$. 

As they move along the track, the particles assemble to form dynamic clusters mediated by hydrodynamic, magnetic dipole-dipole, and excluded volume interactions. The  clusters of different shapes and sizes move at different speeds within the common driving field. Fast-moving clusters ($U > R \Omega$) roll into the outer wall of the track thereby disrupting their organization and slowing their motion. These clusters form a moving front or ``stampede'' of rollers that orbit the track at a frequency $\Omega$ specified by the driving field (Figure \ref{fig:rollers}). By contrast, slow-moving clusters ($U < R\Omega$) that fail to keep up with the stampede move erratically in place until the moving front of rollers returns. As the stampede passes over the slow-moving clusters, the latter may be incorporated into the stampede or otherwise reconfigured by interactions with other clusters.  The motivating hypothesis of this system is that over time the particles will organize to form persistent clusters distinguished by their stability in the driving field. 

To test this hypothesis, we analyze microscopy videos of particle motion to estimate the mutual information $\hat{I}(\tau)$ as a function of lag time (Figure \ref{fig:rollers}c). For each time interval $\tau$, the videos are coarse grained to create a sequence of $n$ binary images containing $h\times w$ pixels comparable in size to the individual particles. Following Cavagna \emph{et al.},\autocite{cavagna2020vicsek} we use $z$-order (Morton space-filling) curves to transform the three-dimensional ($n\times h\times w$) bit array into a one-dimensional bit string. These orderings preserve locality of the bits in space and time and are therefore better in capturing time correlations in cluster dynamics.  We use the LZ78 algorithm to compress the bit string and estimate its binary code length $\mathcal{L}$ using the bound in equation (\ref{eq:bound}). Dividing by the number of images $n$, we obtain the CID measuring the number of bits of information per frame in the coarse grained video. To estimate the amount of heritable information $\hat{I}(\tau)$, we repeat this analysis on shuffled image sequences and evaluate the difference in the respective CIDs following equation (\ref{eq:Icid}).

As expected, the heritable information $\hat{I}(\tau)$ decreases monotonically with increasing lag time as physical memories of the system's past fade. Interestingly, however, these memories decay slowly as approximated by a truncated power-law with an exponent of $\alpha\approx 0.5$ over the lag times investigated. At longer times, the heritable information $\hat{I}(\tau)$ decays exponentially with a rate comparable to the slow rotation frequency $\Omega$ of the driving field.  Longer experiments are required to  determine whether or not any information is preserved for longer times, $\Omega \tau > 1$.  The presence of such long-lived correlations---if they exist---would be suggestive of the type of persistent patterns from which the physicochemical precursors of life evolved. Despite the lack of evidence for such patterns here, the proposed methodology should be useful for detecting dynamical persistence in this and other experimental models of active matter.\autocite{marchetti2013hydrodynamics, bechinger2016active}  

CID-based measures of heritable information can be  applied in practice using relatively small numbers of sampled states. In the present example, the space of binary image sequences contains an astronomical $2^{h w n}$ possible states; however, meaningful estimates of heritable information are obtained from a single image sequence of only $n\sim100$ frames.  Due to the physical principles of continuity and locality, the dynamics on this large state space admit a far more concise description based on moving particles and their local interactions. Consequently, the use of the z-order curve during compression is essential to capturing the decomposition of the system into local subsystems exhibiting similar dynamics. Almost no heritable information is detected when the image sequences are encoded differently prior to compression. Presumably, the dependence of $\hat{I}$ on the details of state encoding should vanish in the limit  as $n\rightarrow\infty$; however, this asymptotic behavior has limited relevance in practice. Care is therefore required in interpreting $\hat{I}$ for finite $n$. Our present analysis captures local forms of heredity (e.g., the persistence of a particular cluster geometry) but is unlikely to capture more complex forms based on communication and cooperation among clusters, which require more data to resolve.
 
We emphasize that the use of universal compression algorithms allows for the detection of persistent patterns with little to no knowledge of the patterns themselves or the mechanisms by which they persist. Such patterns are distinguished by surprising amounts of heritable information that persist over time scales much longer than those of internal relaxation or external driving. Once discovered, the relevant patterns that enable compression can be interrogated more explicitly to identify the persistent features of the system and understand their dynamical stability.  By searching for information about the past, one aims to discover new modes of organization that enable that information to survive the passage of time.  We hypothesize that this perspective---one that elevates heritable information above other features of life---will help in identifying experimental models that exhibit primitive forms of heredity relevant to life's origins.

%%%%%%%%%%%%%%%%%%%%%%%%%%%%%%%%%%%%%%%%%%%%%%%%%%%%%%%%%%%%%%%
%%%%%%%%%%%%%%%%%%%%%%%%%%%%%%%%%%%%%%%%%%%%%%%%%%%%%%%%%%%%%%%
\section*{Conclusions}

So what kind of explanation can we expect to find for the origins of life? The answer advocated here is a conceptual framework based on heritable information in dynamical systems. Within this framework, we consider an ensemble of possible dynamical systems. The prior probability assigned to each possibility is chosen to reflect our knowledge of the physical world and the constraints it imposes (e.g., locality, continuity). We imagine that many of these hypothesized systems occur somewhere in the world like so many primordial soups with different components and environments. As the systems change in time, we look for information about their past within their present configurations. For the majority of such systems, the present state is explained by their components and environment alone. Others, however, contain additional information about their history.  In rare instances, these histories are not frozen relics of the past but are actively maintained by dissipative processes that appear organized for the purpose of preserving heritable information.  Once identified, these persistent patterns capture our attention and demand explanation of the mechanisms---the so-called ``free floating rationales'' \autocite{dennett2017bacteria}---that underlie their existence. The origins of life is informed by this persistence filter whereby patterns of organization are selected by the passage of time on the basis of their persistence alone. In these concluding remarks, we highlight some of the (many) remaining questions about this proposed framework.

% Static vs Dynamical Persistence 
In our numerical investigation, we imposed the artificial constraint that all microstates have a common lifetime. As a result, the persistent systems we identify are necessarily dissipative and require purposeful organization to preserve information for times longer than this microscopic time scale. By contrast, real systems have a spectrum of microstate lifetimes, some of which may be extremely long as evidenced by the stability of the agate (Figure \ref{fig:intro}a). Our measure of persistence based on mutual information between the past and present does not discriminate between these different types of persistence. Further work is needed to more clearly distinguish dynamic and static persistence---that of the bacterium and the agate, respectively (Figure \ref{fig:intro}). In the present context, we use the term dynamical persistence to describe systems that preserve heritable information on timescales longer than even the longest microstate lifetime. Such persistence is not a property of the microstates themselves but rather their mesoscopic organization. The connection between material organization and heritable information is a central theme of this Chapter and demands further investigation. As Darwinian evolution informs our understanding of biological function, heritable information may provide new principles for understanding material organization in nonequilibrium systems.

% Multiple Levels, Fluctuating Environments 
We further impose the simplifying assumption that each dynamical system is isolated within its own unchanging environment. In reality, these systems are embedded within larger systems and interact with one another to create complex, fluctuating environments. The carving of the world into system and environment, microstates and mesostates, fast and slow processes is messy business filled with ambiguities. The simple models described here are useful for building intuition precisely because they are free from such ambiguities. The study of experimental systems---even those as simple as the colloidal rollers discussed above---requires difficult choices to best ``carve nature at its joints''. In navigating these choices, the concept of heritable information can provide a useful guide. We seek the most concise representations that capture the system's heritable information on a prescribed time scale.  For this reason, we coarse grain the images of the rollers as to omit features smaller than the particles themselves.  Such \emph{ad hoc} coarse graining procedures can surely be improved using machine learning (ML) techniques such as variational autoencoders, which create coarse grained representations that maximize a specified objective function (e.g., the amount of heritable information). The pattern-finding capabilities of ML algorithms offer many exciting opportunities to identity persistent patterns of heritable information in diverse material systems.

% who's information? 
One question that we have not attempted to address is the semantic meaning of heritable information. We instead adopt Shannon's perspective and analyze dynamical systems as communications channels that transmit information between the past and the present without regard to its meaning. Such channels are a critical prerequisite for the hereditary processes underlying life as we know it. In this Chapter, we have sought to explain how inanimate material systems are capable of transmitting information through time. For example, the 12-state system of Figure \ref{fig:Example} transmits 1 bit of information about its distant past to the present; however, we have not discussed how this bit was written or what it might mean.  Returning to the analogy with Natural Selection, it would appear that persistent patterns succeed in preserving heritable information but lack the capacity for mutation and thereby evolutionary change. This view is incorrect: change is ubiquitous, persistence in spite of such change is rare and the defining characteristic of life.  The distinction between change and persistence is often a matter of time scales.  What persists on one time scale changes over longer times.  Our genome is stable over our lifetime but changes in part over generations even as the vast majority of distinctly human genetic patterns persist longer still. By exploring the mechanisms of dynamical persistence in material systems and estimating their respective probabilities under different conditions, we move closer to understanding the origins of heredity and thereby life itself.

%To understand life's origins, we must explain how inanimate material systems transmit information through time.

% Initial conditions, ergodicity, 
%Another challenge posed by our definition of heritable information $I(\tau)$ is its reliance on expectations over states weighted by their stationary probabilities $\pi_i$. The Markov processes we describe are ergodic: given sufficient time, they sample all states of the system. By contrast, it is argued that life is inherently non-ergodic: the space of possibilities is vastly greater than that of actualities.\autocite{kauffman2019world} How can idealized concepts such as stationary distributions be relevant to quantifying heritable information? 

%It describes all possible histories   
%Expected uncertainty vs. actual uncertainty
%Non-ergodic
%Branching Histories.

%Different initial conditions vs. different systems

%We anticipate that the growing field of thermodynamic inference\kbnote{REF} will contribute new insights into the fundamental relationships between stochastic thermodynamics and dynamical persistence.

%\paragraph{Physical Constraints.}  Possible dynamics are constrained by features of the physical world. These constraints influence whether persistence is inevitable or impossible.

%%%%%%%%%%%%%%%%%%%%%%%%%%%%%%%%%%%%%%%%%%%%%%%%%%%%%%%%%%%%%%%
%%%%%%%%%%%%%%%%%%%%%%%%%%%%%%%%%%%%%%%%%%%%%%%%%%%%%%%%%%%%%%%
\section*{Methods}

%%%%%%%%%%%%%%%%%%%%%%%%%%%%%%%%%%%%%%%%%%%%%%%%%%%%%%%%%%%%%%%
\subsection*{Coarse Graining}

\begin{figure}[t]
    \centering
    \includegraphics{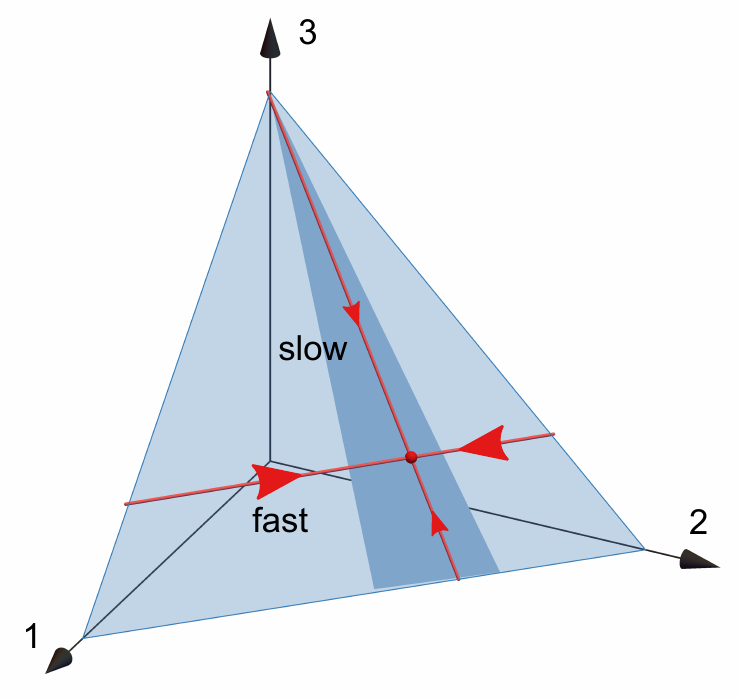}
    \caption{Geometric representation of the 2D probability space (light blue) of a 3-state system. In time, the simplex of possible distributions (dark blue) contracts toward the stationary state (red dot). This system is characterized by time scale separation between its fast eigenmode ($\lambda_3=1.5$) and its slow eigenmode ($\lambda_2=0.013$). Consequently, the system admits a coarse grained representation---here, one in which states 1 and 2 are combined to form a mesostate.}
    \label{fig:CoarseGraining}
\end{figure}

For a system with rate matrix $W$ characterized by $M$ slow eigenvalues, the $N$ states can be clustered into $M$ mesostates to provide a coarse grained description of the system dynamics.  We use a spectral clustering approach to associate each state $i$ with a mesostate $k$ to determine the elements $\Pi_{ki}$ of the conditional probability matrix introduced in the main text.  Two states $i$ and $j$ are considered to be in the same mesostate if they evolve in time to neighboring points in probability space after the relaxation of the fast modes.  The space of possible distributions $p_i$ is constraine by the requirement that $0\leq p_i\leq 1$ and  by the normalization condition $\sum_i p_i = 1$.  As illustrated in Figure \ref{fig:CoarseGraining} for $N=3$, this space can be viewed as an $(N-1)$-dimensional simplex in $N$-dimensional Euclidean space, where each vertex corresponds to a state of the system.  Under the dynamics of the master equation (\ref{eq:master}), the simplex is squeezed, sheared, and rotated by the linear transformation $T(t)$, ultimately collapsing to a single point describing the stationary distribution.  At all times, the collapsing simplex must remain within its initial bounds, thereby restricting the possible eigenvalues and eigenvectors. 

For systems with timescale separation, the relaxation of the fast modes causes the $N$ vertices of the simplex to cluster into $M$ mesostates, which approximate a new simplex of reduced dimensionality. The matrix element $T_{ij}(t)$ describes the $i^{\mathrm{th}}$ coordinate of the $j^{\mathrm{th}}$ vertex at time $t$.  At intermediate times---much longer than the fast time $\lambda^{-1}_{\fast}$ but much shorter than the slow time $\lambda^{-1}_{\slow}$---these vertex coordinates are well approximated as
\begin{equation}
    T_{ij}(t) \approx \sum_{m} Q_{im} Q^{-1}_{mj} 
\end{equation}
where the sum includes only the slow modes. We use the $k$-means clustering algorithm to group these points into $M$ mesostates.  Alternatively, one may cluster points in the ($M-1$)-dimensional eigenspace by applying the $k$-means algorithm to the $N$ coordinates, $Q^{-1}_{mj}$.

%%%%%%%%%%%%%%%%%%%%%%%%%%%%%%%%%%%%%%%%%%%%%%%%%%%%%%%%%%%%%%%
\subsection*{Monte Carlo Optimization}
We use the Metropolis–Hastings algorithm to sample the space of possible dynamics in search of those that minimize some property $U$---for example, the relaxation rate $\RE(\lambda_2)$.  This procedure is analogous to finding the minimum energy configuration of an equilibrium system. The space of possible dynamics corresponds to that of possible adjacency matrices for strongly connected graphs. Starting from one such strongly connected graph, we consider two possible Monte Carlo moves---bit flips and bit swaps---at each step of the Markov chain. During a bit flip, we select one of the $N(N-1)$ possible edges at random and flip the corresponding value of the adjacency matrix $A_{ij}$ to add or remove the edge. During a bit swap, we select at random two elements of the adjacency matrix such that $A_{ij}=1$ and $A_{kl}=0$ and swap their values. If an attempted move breaks the connectivity constraint, the attempt is rejected. Otherwise, the move from the current dynamics $n$ to the candidate dynamics $m$ is accepted with probability $\min(1,e^{-\beta (U_{m}-U_{n})})$.  This algorithm samples the dynamics $n$ in accordance with the Boltzmann distribution $p_n\propto e^{-\beta U_n}$, where $\beta \in [0,\infty)$ is a tunable parameter. To identify the dynamics that minimize $U$, we use simulated annealing\autocite{kirkpatrick1983optimization} to gradually increase $\beta$ thereby biasing the Markov chain towards dynamics with smaller $U$. 

%%%%%%%%%%%%%%%%%%%%%%%%%%%%%%%%%%%%%%%%%%%%%%%%%%%%%%%%%%%%%%%
%\subsection*{Wang-Landau Algorithm\autocite{Wang2001}}
%To determine the probability of finding a network with a certain eigenvalue we use the Wang Landau algorithm. Similar to the MCMC simulations described, above the algorithm searches the space of possible adjacency matrices by follwoing the moves described above. This algorithm uses bins for the target function ($\lambda_2$), and accepts or rejects moves based on the number of times a bin has been visited. Using this procedure the number of times a bin is visited is related to the probability of choosing a network from that bin at random. The minimum eigenvalue for a set of parameters is first determined by MCMC as described above. An estimate for the maximum eigenvalue is determined using $\lambda^{max}=N/(N-1)$. Bins are then setup to be evenly spaced in log space. An extra bin is also created for networks that break the connectivity constraints, this bin is not shown in our results (Figures \ref{fig:WL_scale} and \ref{fig:constraints}).

%%%%%%%%%%%%%%%%%%%%%%%%%%%%%%%%%%%%%%%%%%%%%%%%%%%%%%%%%%%%%%%
%%%%%%%%%%%%%%%%%%%%%%%%%%%%%%%%%%%%%%%%%%%%%%%%%%%%%%%%%%%%%%%
\subsection*{Experiments on Ferromagnetic Rollers}

We prepared magnetic Janus particles (MJPs) by e-beam deposition of successive layers of metal---5 nm Ti adhesive layer, 25 nm Ni magnetic layer---onto monolayers of 4 $\mu$m fluorescent sulfonated polystyrene (PS) particles.\autocite{fei2020magneto} The nickel layer gave the particles a permanent magnetic moment $m\approx 3\times10^{-14}$ A m$^2$ directed parallel to the Janus equator.\autocite{fei2018magneto} To mitigate their agglomeration in water, the Janus particles were coated with a thin layer of silicon dioxide by sputter deposition. The circular track was fabricated in polydimethyl siloxane (PDMS) by soft lithography and treated with a surfactant solution (0.2 w/v\% Pluronic F127 in water) to further prevent particle adhesion. The aqueous dispersion of MJPs were pipetted into the PDMS track and sealed with a glass cover slide using UV-curable adhesive.

During the rolling experiments, the particles settled under gravity to the glass substrate where they were imaged from below by an inverted microscope. The time-varying magnetic field $\ve B(t)$ was generated using a custom triaxial electromagnet \autocite{dou2020programmable} with a waveform given by equation (\ref{eq:B}). The magnitude of the field was held constant with $B_0 = 3$ mT as measured by a gaussmeter. The rolling frequency $\omega$ was chosen to be sufficiently slow such that magnetic particles could role in sync with the applied field---that is, $\omega < m B_0 / \lambda $ where $\lambda\approx8\pi\eta a^3$ is the hydrodynamic resistance to rotation for a sphere in a fluid of viscosity $\eta$. The rotation frequency $\Omega$ on which the direction of the rotating field changes was chosen empirically to permit ``fast'' clusters to follow the field around track but not ``slow'' clusters and individual particles.  Here, we use a rolling frequency of $\omega = 62.8$ rad/s and a rotation frequency of $\Omega=0.0140$ rad/s. The frame rate of image acquisition was $0.67$ fps.

The video analyzed in Figure \ref{fig:rollers} contained 2801 grayscale images with $4908 \times 3264$ pixels each. Each image was cropped, downsampled, and thresholded to create a $320 \times 320$ binary image, where each square pixel ($5.8 \times 5.8~\mu$m$^2$) denotes the presence (1) or absence (0) of a particle at that location. Those pixels located outside of the circular track were always zero and therefore omitted from the analysis. In estimating the mutual information $\hat{I}(\tau)$, we considered 33 time lags $\tau$, each an integer multiple of the time step for image acquisition, $\Delta t = 1.5$ s. For each lag, frames were removed from the binary image sequence such that the remaining frames were spaced by $\tau$.  Following Cavagna \emph{et al.},\autocite{cavagna2020vicsek} we ordered the 3D bit arrays into six 1D bit strings using as many permutations of the $z$-order curve (e.g., $(x,y,t)$, $(y,t,x)$, etc.). The resulting CIDs were averaged over these six strings to improve the isotropy of the analysis and avoid privileging some dimensions over others.  Each bit string was compressed using the LZ78 algorithm (Figure \ref{fig:CID}), and the CID computed using equation (\ref{eq:CID}) with the binary code length $\mathcal{L}$ estimated by equation (\ref{eq:bound}) and the number of frames $n$ determined by the lag as $n=\lfloor 2801 \Delta t/\tau \rfloor$. Finally, the binary image sequence was randomly shuffled in time and its CID computed. In accordance with equation (\ref{eq:Icid}), the mutual information estimate $\hat{I}(\tau)$ was evaluated as the difference between the average CID of eight shuffled sequences and that of the original sequence.

%There are a few limitations of the application of this procedure worth noting. First the minimum $\tau$ was limited by the frame rate. To investigate persistence on the order of the rolling frequency $\omega$, a frame rate multiple orders of magnitude would have been needed. Second the the number of frames for large $\tau$ was limited, which likely led to the estimate deviating further from the true conditional entropy. The effect of a variable number of frames should be looked at further in future studies. Third the number of random shuffles used to estimate $\langle \mathrm{CID(x_{\sh})} \rangle$ was limited. Ideally all possible permutations of the frames should be done in order to best estimate the marginal entropy. The analysis procedure should be partially adapted to each specific application keeping in mind these limitations.

%%%%%%%%%%%%%%%%%%%%%%%%%%%%%%%%%%%%%%%%%%%%%%%%%%%%%%%%%%%%%%%
%%%%%%%%%%%%%%%%%%%%%%%%%%%%%%%%%%%%%%%%%%%%%%%%%%%%%%%%%%%%%%%
\printbibliography

%%%%%%%%%%%%%%%%%%%%%%%%%%%%%%%%%%%%%%%%%%%%%%%%
%%%%%%%%%%%%%%%%%%%%%%%%%%%%%%%%%%%%%%%%%%%%%%%%
%%%%%%%%%%%%%%%%%%%%%%%%%%%%%%%%%%%%%%%%%%%%%%%%
\section*{Acknowledgements}

This work was supported in part by the Center for Bio-Inspired Energy Science, an Energy Frontier Research Center funded by the U.S. Department of Energy, Office of Science, Basic Energy Sciences under Award DE-SC0000989.  K.J.M.B. is grateful to L.~Bishop for useful discussions of this work.

\appendix
\section{Persistence in Equilibrium Systems}
\label{app:equil}

Equilibrium systems with two mesostates that minimize the second eigenvalue $\lambda_2$ are organized into two clusters connected by a linear chain of states (Figure \ref{fig:equilibrium}b). Each cluster contains $N_c$ states, which are fully connected to one another by reversible transitions; the linear chain contains $N_l$ states. Here, we aim to estimate the minimum eigenvalue $\lambda_2^{\min}$ and the allocation of the $N=2N_c+N_l$ states between the clusters and the chain.

To simplify the analysis, we coarse grain each cluster as a single state with an escape rate $k$. The probability of being in any state within the cluster relaxes quickly to $N_c^{-1}$ on time scales of order one. The rate of exiting the cluster from the single outlet is also $N_c^{-1}$ owing to our rules relating transition rates to the number of outgoing transitions.  Therefore, we can approximate the escape rate for the coarse grained cluster as $k=N_c^{-2}$. 

Within the linear chain of $N_l$ states, the probability of being in state $i$ evolves as 
\begin{equation}
    \frac{d{p}_i}{d t} = \tfrac{1}{2}p_{i-1} - p_i + \tfrac{1}{2}p_{i+1} \label{eq:eq1}
\end{equation}
for the internal states $i=2,\dots,N_l-1$. At the ends of the chain, the state probabilities evolve as  
\begin{align}
    \frac{d p_1}{d t} &= k p_L - p_1 + \tfrac{1}{2}p_{2}
    \\
    \frac{d p_{N_l}}{d t} &= \tfrac{1}{2} p_{N_l-1} - p_{N_l} + k p_R
\end{align}
where $p_L$ and $p_R$ are the probabilities that the system is in the coarse grained clusters on the left and right, respectively.  These probabilities evolve in time as 
\begin{align}
    \frac{d p_L}{d t} &= -k p_L + \tfrac{1}{2}p_1
    \\
    \frac{d p_R}{d t} &= \tfrac{1}{2} p_{N_l} - k p_R \label{eq:eq5}
\end{align}
Together, equations (\ref{eq:eq1})--(\ref{eq:eq5}) specify the dynamics of the system neglecting the internal dynamics within the two clusters, which are assumed to be fast.

For large numbers of states, we can approximate the discrete states $i=1,2,\dots,N_l$ by a continuum $x \in [0,N_l]$ where the probability density $p(x,t)$ is governed by the diffusion equation
\begin{equation}
    \frac{\partial p}{\partial t} = \frac{1}{2} \frac{\partial^2 p}{\partial x^2}
\end{equation}
At the boundaries ($x=0,N_l$), conservation of probability requires that 
\begin{align}
    \frac{d p_L}{d t} &=\left. \frac{1}{2}\frac{\partial p}{\partial x}\right\rvert_{0} = -k p_L + \frac{1}{2}p(0,t) \label{eq:pl}
    \\
    \frac{d p_R}{d t} &=\left. -\frac{1}{2}\frac{\partial p}{\partial x}\right\rvert_{N_l} = \frac{1}{2}p(N_l,t) -k p_R \label{eq:pr}
\end{align}
The discrete equations above correspond to a finite difference approximation of this continuum description.  
We seek the slowest decaying eigenmode of the form $p(x,t)= \hat{p}(x) e^{-\lambda t}$. Substituting this expression into the diffusion equation, we obtain harmonic solutions $\hat{p}(x)$ of the form 
\begin{equation}
    \hat{p}(x) = A \cos(\sqrt{2\lambda} x) + B \cos(\sqrt{2\lambda} x)
\end{equation}
Substituting this result into equations (\ref{eq:pl}) and (\ref{eq:pr}) we obtain the following characteristic equation for $\lambda$
\begin{equation}
    \sqrt{2\lambda} \left(2 (k-\lambda )^2-\lambda \right) \sin(\sqrt{2 \lambda } N_l) - 4 \lambda (\lambda -k) \cos(\sqrt{2\lambda} N_l) = 0
\end{equation}
Anticipating that $\lambda\propto N^{-3}$, we can expand this equation in a Taylor series about $\lambda=0$ to obtain a quadratic approximation for the first two roots valid for large numbers of states $N\gg1$.
%\begin{equation}
%    0 = \lambda\left(\lambda - \frac{6 k (k N_l+1)}{N_l (2 k (N_l (k N_l+3)+6)+3)+6}\right)
%\end{equation}
Substituting $k=N_c^{-2}$ and introducing $\alpha = N_l / N$, the second root $\lambda_2$ can then be approximated as 
\begin{equation}
    \lambda_2 = \frac{8}{ \alpha (\alpha -1)^2 N^3} + O(N^{-4})
\end{equation}
which reaches it minimum value of $\lambda^{\min}_2 = 54/N^{3}$ for $\alpha=1/3$.  Consistent with our numerical results (Figure \ref{fig:equilibrium}b), the states are partitioned equally in three groups corresponding to the two clusters and the linear chain.  The minimum eigenvalue $\lambda_2^{\min}$ scales as $N^{-3}$ as illustrated in Figure \ref{fig:equilibrium}a.  The difference in the prefactor (54 here vs.~47 in Figure \ref{fig:equilibrium}a) is due to the coarse graining of the connected clusters in our present analysis.

%%%%%%%%%%%%%%%%%%%%%%%%%%%%%%%%%%%%%%%%%%%%%%%%%%%%%%%%%%%%%%%
%%%%%%%%%%%%%%%%%%%%%%%%%%%%%%%%%%%%%%%%%%%%%%%%%%%%%%%%%%%%%%%
\end{document}